\documentclass[twocolumn,english,american,pra]{revtex4-2}
\usepackage[LGR,T1]{fontenc}
\usepackage{textcomp}
\usepackage[utf8]{inputenc}
\usepackage{color}
\usepackage{babel}
\usepackage{amsmath}
\usepackage{amssymb}
\usepackage{graphicx}
\usepackage[]
 {hyperref}

\makeatletter

\DeclareRobustCommand{\greektext}{%
  \fontencoding{LGR}\selectfont\def\encodingdefault{LGR}}
\DeclareRobustCommand{\textgreek}[1]{\leavevmode{\greektext #1}}

\providecommand{\tabularnewline}{\\}

\usepackage{tikz}
\usetikzlibrary{quantikz2}
\usepackage{diagbox}
\usepackage{orcidlink}

\makeatother

\begin{document}
\selectlanguage{english}%
\global\long\def\ket#1{|#1\rangle}%

\global\long\def\bra#1{\langle#1|}%

\global\long\def\braket#1#2{\langle#1|#2\rangle}%

\global\long\def\tr#1{\text{tr}#1}%

\title{Digital quantum simulation of squeezed states via enhanced bosonic
encoding in a superconducting quantum processor}
\author{Hengyue Li \orcidlink{0009-0002-6903-2994}}
\email{lihy@arclightquantum.com}

\affiliation{Arclight Quantum Computing Inc., Beijing, 100191, People's Republic
of China}
\author{Yusheng Yang \orcidlink{0000-0002-3584-4954}}
\affiliation{Arclight Quantum Computing Inc., Beijing, 100191, People's Republic
of China}
\author{Zhe-Hui Wang}
\affiliation{QuantumCTek (Shanghai) Co., Ltd., Shanghai, 200120, People's Republic
of China}
\author{Shuxin Xie}
\affiliation{QuantumCTek (Shanghai) Co., Ltd., Shanghai, 200120, People's Republic
of China}
\author{Zilong Zha}
\affiliation{China Telecom Quantum Information Technology Group Co., Ltd., Hefei,
230031, People's Republic of China}
\author{Hantao Sun}
\affiliation{China Telecom Quantum Information Technology Group Co., Ltd., Hefei,
230031, People's Republic of China}
\author{Jie Chen}
\affiliation{China Telecom Quantum Information Technology Group Co., Ltd., Hefei,
230031, People's Republic of China}
\author{Jian Sun}
\affiliation{Arclight Quantum Computing Inc., Beijing, 100191, People's Republic
of China}
\author{Shenggang Ying}
\email{yingsg@ios.ac.cn}

\affiliation{Institute of Software, Chinese Academy of Sciences, Beijing, 100190,
People's Republic of China}
\affiliation{Arclight Quantum Computing Inc., Beijing, 100191, People's Republic
of China}
\begin{abstract}
We present a fully digital approach for simulating single-mode squeezed
states using an enhanced bosonic encoding strategy on a circuit model,
and demonstrate it on a superconducting quantum processor through
a cloud platform. By mapping up to $2^{n}$ photonic Fock states onto
$n$ qubits, our framework leverages Gray-code-based encodings to
reduce gate overhead compared to conventional one-hot or binary mappings.
We further optimize resource usage by restricting the simulation to
Fock states with even numbers of photons only, effectively doubling
the range of photon numbers that can be represented for a given number
of qubits. To overcome noise and finite coherence in current hardware,
we employ a variational quantum simulation protocol, which adapts
shallow, parameterized circuits through iterative optimization. Implemented
on the \textit{Zuchongzhi-2} superconducting platform, our method
demonstrates squeezed-state dynamics across a parameter sweep from
vacuum state preparation ($r=0$) to squeezing levels exceeding the
Fock-space truncation limit ($r>1.63$). Results of demonstration,
corroborated by quantum state tomography and Wigner-function analysis,
confirm high-fidelity state preparation and demonstrate the potential
of Gray-code-inspired techniques for realizing continuous-variable
physics on near-term, qubit-based quantum processors.
\end{abstract}
\maketitle

\section{Introduction}

With continued development of quantum circuit-based architectures
\cite{Nielsen2010}, these systems are poised to complement specialized
quantum machines in addressing applications ranging from materials
simulation \cite{Stanisic2022,Li_2024} and quantum phase simulations
\cite{PhysRevA.111.022618,Hu2025} to cryptographic protocols \cite{365700,Shor2}
and pharmaceutical research \cite{8585034}. In pursuit of a universal
quantum processor, a variety of hardware platforms---ranging from
superconducting circuits \cite{SC1,SC_review} and trapped ions \cite{ion1,ion2}
to neutral atoms \cite{neutral1,neutral2} and photonic systems \cite{KLM,MBQC}---are
under intensive development. Each implements quantum logic within
the circuit model, whereby qubits undergo sequences of well-controlled
gate operations. Superconducting qubits, for instance, have made rapid
strides in qubit count and fidelity, aided by improvements in fabrication
and control electronics. Trapped-ion platforms likewise provide high
qubit connectivity and long coherence times, while neutral-atom arrays
offer scalability through optical trapping techniques.

Intriguingly, photonic quantum computing has evolved along two main
pathways: the discrete-variable (DV) approach \cite{FBQC1} and the
continuous-variable (CV) approach \cite{Xanadu1}. DV photonics treats
single photons as qubits for a circuit-model calculation, but suffers
from non-deterministic entangling gates \cite{KLM}. In response,
measurement-based quantum computing (MBQC) \cite{MBQC} protocols
were devised, leveraging percolation \cite{percolation} techniques
to build large cluster states that enable universal DV quantum logic.
Conversely, CV photonics encodes information in the infinite-dimensional
Hilbert space of electromagnetic modes---often referred to as the
“phase space.” This method also supports universal quantum computation,
for example through the Gottesman--Kitaev--Preskill (GKP) encoding
\cite{GKP}, which can implement a circuit model using CV states.
Given these capabilities, it is both natural and beneficial to explore
the inverse scenario---namely, simulating CV photonic (or more broadly,
bosonic) physics on digital, qubit-based hardware. By mapping continuous-variable
systems onto qubit registers, one can harness existing circuit-model
devices to investigate and emulate complex quantum processes characteristic
of photonic or vibrational modes.

To simulate CV systems on qubit-based quantum hardware, photonic Fock
states must be explicitly mapped to multi-qubit states via encoding.
A commonly considered direct method is one-hot (unary) encoding \cite{onehot1},
where each Fock state $\ket n_{F}$ corresponds to a computational
basis state $\ket{0...010...0}$. While intuitive, this approach incurs
linear resource overhead: representing $d$ photon levels requires
$d+1$ qubits, making it impractical for large $d$. To mitigate this,
the Gray code (also known as the \textit{reflected binary code}) has
been studied \cite{graycode1}. Unlike one-hot encoding, which uses
only $n$ of the available $2^{n}$ basis states, Gray codes fully
exploit the Hilbert space, reducing qubit counts. Studies comparing
various encoding methods \cite{Sawaya2020,Kottmann2021} reveal that
the optimal choice depends critically on the target application. Recent
work has also identified novel approaches such as encoding bosons
via fermions followed by mapping fermionic states to qubits through
the Jordan-Wigner transformation \cite{Chin2024}, though this method
introduces trade-offs in resource allocation and gate complexity.

In this work, we present a generalized encoding framework that maps
$2^{n}$ photonic states onto $n$ qubits. We analyze the efficiency
of these encodings and demonstrate that the Gray code---a member
of this encoding family---achieves high efficiency in bosonic system
simulations, for which we provide theoretical justification. Furthermore,
we implement a quantum simulation of a bosonic system on real hardware.
Building on foundational work demonstrating two-photon squeezed state
simulations with one-hot encoding \cite{squeeze1}, our protocol
extends the accessible photon number to $6$ through optimized bosonic
mapping. This advancement enables exploration of squeezing parameters
up to $r=2$, operating beyond the regime where the truncated Fock-space
approximation breaks down due to saturation of maximum photon state
population. To address the challenges of noise and limited coherence
in quantum hardware, we implement a variational quantum simulation
(VQS) protocol \cite{VQS1,VQS2,VQS3}. This approach leverages parameterized
quantum circuits optimized through hybrid quantum-classical feedback,
adaptively balancing circuit depth and simulation accuracy. By avoiding
the strict gate sequence requirements of Suzuki-Trotter decomposition
\cite{Suzuki}, the VQS framework enables robust evolution of squeezed
states under realistic device.

The remainder of this paper is organized as follows. In Section \ref{secEncode},
we introduce our generalized encoding framework for mapping up to
$2^{n}$ photonic Fock states onto $n$ qubits. Section \ref{subsecGE}
outlines the construction of multi-qubit representations of the bosonic
ladder operators and explains how to realize these encodings in practice.
Section \ref{subsecGC} then discusses why Gray-code-based encodings
are especially efficient for most bosonic simulations. In Section
\ref{secSSS}, we demonstrate our method by simulating single-mode
squeezing on a superconducting quantum processor (Zuchongzhi-2) \cite{guodun_yingjian}
provided by \textit{QuantumCTek Co., Ltd.} Section \ref{subsec:EH}
describes a specialized variant of the Gray code that encodes only
even-photon Fock states, effectively doubling the highest photon number
that can be represented. Section \ref{subsec:VQS} presents our variational
quantum simulation (VQS) procedure, designed to reduce circuit depth
while maintaining accuracy. In Section \ref{sec:results}, we compare
the simulated results to exact theoretical benchmarks and analyze
the hardware performance. Finally, Section V summarizes our main conclusions.

\section{Encoding bosonic operator}\label{secEncode}

A general bosonic Hamiltonian is expressed as $H=H\left(\{b_{i}^{\dagger},b_{j}\}\right)$,
where $b_{i}$ ($b_{i}^{\dagger}$) is the annihilation (creation)
operator for the $i$-th bosonic mode. These operators satisfy the
commutation relation $[b_{i},b_{j}^{\dagger}]=\delta_{ij},$ where
$\delta_{ij}$ is the Kronecker delta function. The entire Hilbert
space of the system, $\mathbb{H}$ , which contains $m$ modes, is
a direct product of the subspaces of each mode, $\mathcal{H}_{i}$,
and is given by 
\begin{align}
\mathcal{\mathbb{H}} & =\mathcal{H}_{1}\otimes\mathcal{H}_{2}\otimes...\mathcal{H}_{m}.
\end{align}
Our objective is to encode the single-mode space $\mathcal{H}$ onto
a circuit model. The extension to multiple modes is straightforward
by employing a larger quantum circuit with additional qubits. The
single-mode Hilbert space $\mathcal{H}$ is spanned by the Fock basis
$\{\ket i_{F}\}$, where $i$ are non-negative integers, and the basis
state $\ket i_{F}$ represents a system containing $i$ bosons (e.g.,
photons). The subscript \textquotedbl$F$\textquotedbl{} distinguishes
the Fock states from the computational basis (denoted without subscripts)
used to encode the Fock states. For practical simulations, a cutoff
is introduced at a maximum boson number, $N_{\text{Max}}$, such that
$\ket{N_{\text{Max}}}$ is the highest allowed state, satisfying $b^{\dagger}\ket{N_{\text{Max}}}_{F}=0$.

\subsection{General encoding}\label{subsecGE}

In this paper, we focus on encoding $N=2^{n}$ Fock states using $n$
qubits. Specifically, we encode a Fock-space of dimension $N$ by
mapping each Fock state to a unique $n$-qubit basis state. A code
$c=\{c_{0},c_{1},...,c_{N-1}\}$ represents a permutation of the natural
sequence $(0,1,...,N-1)$, resulting in $N!$ possible encoding schemes,
including the natural encoding (also known as the binary code \cite{Sawaya2020}). 

To systematically organize these encoding schemes, we arrange all
possible codes in dictionary order, forming a set $\mathcal{C}_{n}$.
In this ordering, sequences are compared by examining each position
sequentially; the sequence with the smaller value at the first differing
position is considered smaller. Thus, the set is denoted as: $\mathcal{C}_{n}=\{\mathcal{C}_{n}^{[0]},\mathcal{C}_{n}^{[1]},...,\mathcal{C}_{n}^{[N!-1]}\}$
where $\mathcal{C}_{n}^{0}$ corresponds to the natural (binary) encoding.

Any code $c\in\mathcal{C}_{n}$ can be used to encode a single bosonic
mode by representing the Fock state $\ket i_{F}$ with the computational
basis state $\ket{c_{i}}$.  The annihilation operator is encoded
as:
\begin{align}
b & =\sum_{i=1}^{N-1}\sqrt{i}\mathcal{X}_{i,i-1}\mathcal{P}_{i},\label{eq1}
\end{align}
where $\mathcal{X}_{i,i-1}$ is a tensor product of Pauli $X$ operators
acting on the qubits where $c_{i}$ and $c_{i-1}$ differ, and $\mathcal{P}_{i}=\ket{c_{i}}\bra{c_{i}}$
is a projector onto the state $\ket{c_{i}}$ and given by 
\begin{align}
\mathcal{P}_{i} & =\frac{1}{2^{n}}\sum_{\alpha=0}^{N-1}(-1)^{W(i\&\alpha)}\otimes_{k=0}^{n-1}Z_{k}^{\alpha_{k}},\label{eq3}
\end{align}
where $\alpha_{k}$ is the bit value on the $k$-th position of the
integer $\alpha$, $W(*)$ denotes the Hamming weight, i.e., the number
of “1”s in the binary representation, and ``$\&$'' represents the
bitwise AND operation. In this paper, $X_{i}$, $Y_{i}$, and $Z_{i}$
denote the Pauli $X$, $Y$, and $Z$ operators acting on the $i$th
qubit. The detailed derivation of Eq. \ref{eq1} and Eq. \ref{eq3}
is provided in Appendix \ref{appendix:Projector}. 

Here, we provide an example of the encoding using $c=\mathcal{C}_{2}^{[1]}=\{0,1,3,2\}$.
In this case, the working basis $\ket 3$ (which corresponds to $\ket{11}$
in binary representation) represents the two-boson state $\ket 2_{F}$,
and $\ket 2$ (which corresponds to $\ket{10}$ in binary representation)
represents the three-boson state $\ket 3_{F}$. The annihilation operator
is given by:
\begin{align}
b & =\ket 0\bra 1+\sqrt{2}\ket 1\bra 3+\sqrt{3}\ket 3\bra 2.\label{eq2}
\end{align}
For calculating the middle term of Eq. \ref{eq2}, we have: $\mathcal{X}_{2,1}=X_{1}\otimes I_{0}$
since $1$ and $3$ differ at the second qubit (adopting 0-based indexing,
the second index corresponds to $X_{1}$). The projector is given
by $\mathcal{P}_{1}=\frac{1}{4}\left(\boldsymbol{1}-Z_{0}-Z_{1}+Z_{1}\otimes Z_{0}\right).$
The final expression of the encoded operator $b$ is:
\begin{align*}
b & =\frac{1+\sqrt{3}}{4}X_{0}+i\frac{1-\sqrt{3}}{4}Y_{0}+\frac{\sqrt{2}}{4}X_{1}+\frac{\sqrt{2}}{4}iY_{1}\\
 & +\frac{1-\sqrt{3}}{4}Z_{1}X_{0}+i\frac{1+\sqrt{3}}{4}Z_{1}Y_{0}\\
 & -\frac{\sqrt{2}}{4}X_{1}Z_{0}-\frac{\sqrt{2}}{4}iY_{1}Z_{0}
\end{align*}

The expression of the encoded Hamiltonian is obtained by substituting
the encoded $b$ into the formula $H=H(b^{\dagger},b)$. To reduce
the number of terms in the final Hamiltonian expression, the key idea
is to minimize the number of terms in $b$. Let us examine Eq. \ref{eq1}
and Eq. \ref{eq3} more carefully to achieve this reduction.

As shown in Eq. \ref{eq3}, the number of terms in the projector $\mathcal{P}_{i}$
is fixed at $N$. All terms (indexed by different values of $\alpha$)
in $\mathcal{P}_{i}$ are direct products of $Z$ and $I$ operators
only. Moreover, for different $i$, the projectors $\mathcal{P}_{i}$
have identical forms except for different multiplicative factors.
In fact, we have: $\sum_{i}\mathcal{P}_{i}=\boldsymbol{1}.$ 

Next, we discuss the effect of the term $\mathcal{X}_{i,i-1}$. The
operator $b$ represents the \textquotedbl ladder\textquotedbl{}
operation that destroys a boson in the system state. In the first
quantization representation, it is a summation of all possible transitions
between photon states that differ by one photon. Therefore, $\mathcal{X}_{i,i-1}$
must include at least one $X$ operator on a qubit where its product
with $Z$ ($I$) (within the projector $\mathcal{P}_{i}$) generates
$Y$ ($X$) terms. As a result, each term in $\mathcal{X}_{i,i-1}\mathcal{P}_{i}$
for different $i$ has a distinct form. However, it is still possible
to combine some terms. If there is only one $X$ in each term of $\mathcal{X}_{i,i-1}$,
there are only $\left(\begin{array}{c}
n\\
1
\end{array}\right)=n$ different $\mathcal{X}_{i,i-1}$. Therefore, there are at most $nN=n2^{n}$
terms in $b$. If $\mathcal{X}_{i,i-1}$ contains more $X$ operators,
the number of terms in Eq. \ref{eq1} increases. For the case of $n=3$,
we counted the number of terms in $b$ for all codes in $\mathcal{C}_{3}$
and presented the results in Table \ref{tab1}. Out of the total $N!=40320$
codes, the encoded $b$ contains the minimal number of terms $nN=24$
in $4032$ codes. We define a subset $\mathcal{D}_{n}\subset\mathcal{C}_{n}$
that contains codes where there is only one $X$ in each $\mathcal{X}_{i,i-1}$
in Eq. \ref{eq1}. The Hamming distance between neighboring numbers
in the codes in $\mathcal{D}_{n}$ is 1. We counted that there are
144 entries in $\mathcal{D}_{3}$.

\begin{table}
\begin{tabular}{|c|c|}
\hline 
Number of terms in $b$ & Counts codes\tabularnewline
\hline 
\hline 
24 & 4032\tabularnewline
\hline 
32 & 14784\tabularnewline
\hline 
40 & 14784\tabularnewline
\hline 
48 & 6720\tabularnewline
\hline 
\end{tabular}\caption{ Distribution of the number of terms in the encoded annihilation
operator $b$ across different encoding schemes in $\mathcal{C}_{3}$. }\label{tab1}

\end{table}

\subsection{Gray code}\label{subsecGC}

In many previous works, the Gray code is often explained simply as
a code where the Hamming distance between neighboring numbers is $1$,
which is a definition captured by $\mathcal{D}_{n}$ in this paper.
However, here we clarify that the Gray code is just one specific instance
within $\mathcal{D}_{n}$. The Gray code is defined by the recursive
generation formula given by \cite{graycode1} 

\begin{align}
G_{n} & =(\boldsymbol{0}\cdot G_{n-1},\boldsymbol{1}\cdot\bar{G}{}_{n-1}),\label{eq4}
\end{align}
with $G_{1}=(0,1)$, where $\bar{G}_{n-1}$ represents the reverse
order of $G_{n-1}$. The advantage of using the Gray code $G_{n}$
lies in the established rules for generating a specific instance within
the $\mathcal{D}_{n}$ codes. As discussed above, $\mathcal{D}_{n}$
is an excellent option for encoding the operator $b$. These codes
are prime candidates for studying Hamiltonians, such as those involving
an external field $g(b+b^{\dagger})$ or the generator of the displacement
operator $\alpha b^{\dagger}-\alpha^{*}b$.

If the Hamiltonian container higher order terms 
\begin{align}
b^{k} & =\sum_{i=0}^{N-1-k}\sqrt{\frac{(i+k)!}{i!}}\ket i_{F}\bra{i+k}_{F}\text{, }(k\le N-1),\label{eq5}
\end{align}
 we have different strategies. As shown in Eq. \ref{eq5}, in the
summation terms of the expression $b^{k}$, the $i$-th term is coupled
with all terms $i+k$, $i+2k$,..., and thus the entire summation
can be partitioned into $k$ distinct groups. Each group contains
terms that are separated by a multiple of $k$. We rewritten the expression
of $b^{k}$ in a form of summation of different groups as 
\begin{align}
b^{k} & =\sum_{\delta=0}^{k-1}\left[\sum_{i=0}^{I(k,\delta)}\sqrt{\frac{\left(ik+k+\delta\right)!}{(ik+\delta)!}}\ket{ik+\delta}_{F}\bra{(i+1)k+\delta}_{F}\right],\label{eq15}
\end{align}
where $I(k,\delta)=\left\lfloor \frac{1}{k}(N-1-\delta)-1\right\rfloor $.
We define a code to be $k$-fold if, within the encoded sequence,
any two codewords that are separated by $k$ positions have a Hamming
distance of $1$. A $k$-fold encoding is characterized by partitioning
the code into $k$ groups, where each group consists of elements that
are spaced $k$ apart. This grouping ensures that the encoding operator
adheres to the structure defined in Eq. \ref{eq15}.

\begin{figure}
\includegraphics[scale=0.55]{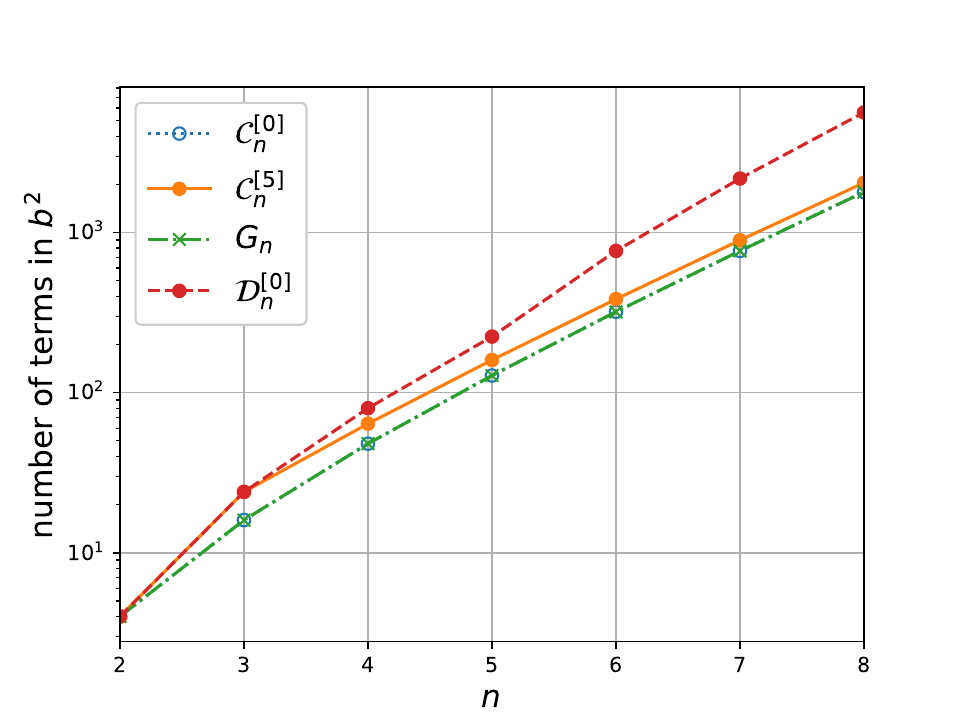}\caption{ Number of terms in the encoding operator $b^{2}$ as a function
of qubit number $n$. The lines represent different codes: $\mathcal{C}_{n}^{[0]}$
(blue, solid with circles), $\mathcal{C}_{n}^{[5]}$ (orange, solid
with circles), $G_{n}$ (green, dashed with crosses), and $D_{n}^{[0]}$
(red, dashed with circles) which is the first code in dictionary order
that ensures a Hamming distance of 1 between adjacent numbers. }\label{fig1}

\end{figure}

As an example, when $k=2$, the summation is divided into two groups:
one representing the sum of the odd-indexed terms and the other representing
the sum of the even-indexed terms, as illustrated by

\begin{align}
b^{2} & =\underbrace{\left(\sqrt{6}\ket 1_{F}\bra 3_{F}+\sqrt{20}\ket 3_{F}\bra 5_{F}+...\right)}_{\text{odd terms}}\nonumber \\
 & +\underbrace{\left(\sqrt{2}\ket 0_{F}\bra 2_{F}+\sqrt{12}\ket 2_{F}\bra 4_{F}+...\right)}_{\text{even terms}}.\label{eq6}
\end{align}
In this scenario, a 2-fold code is naturally suited for encoding $b^{2}$,
as anticipated. As illustrated in Fig. \ref{fig1}, the number of
terms in $b^{2}$ varies with the number of qubits $n$ across several
representative codes. Notably, the binary code $\mathcal{C}_{n}^{[0]}$
is intrinsically a 2-fold code and consistently exhibits the lowest
cost among the compared codes. In contrast, $D_{n}^{[0]}$, which
is the first code in dictionary order that ensures a Hamming distance
of 1 between adjacent numbers, incurs a higher cost, resulting in
a greater number of terms. $C_{n}^{[5]}$ is included for comparison.
As shown in the figure, the Gray code $G_{n}$ also performs favorably,
maintaining a relatively low number of terms.

However, one issue arises: why does the Gray code $G_{n}$ still perform
well, given that $G_{n}\in\mathcal{D}_{n}$ and we believe that 1-fold
codes are not suitable for encoding $b^{2}$? We state that $G_{n}$
is a special code that is also effective for encoding $b^{2}$. To
clarify this, we present an example for the case where $n=3$. The
binary representations of the numbers in $G_{3}$ are regrouped into
even-indexed and odd-indexed terms and are listed in Table \ref{tab2}.
As can be seen, apart from the rightmost qubit, the remaining two
qubits (highlighted in red) still form the Gray code $G_{2}$. In
this configuration, there is a fixed operator $X$ acting on the 0-th
bit position. We will soon see that this additional $X$ does not
adversely affect the encoding process.

\begin{table}
\begin{tabular}{|c|c|}
\hline 
even terms & odd terms\tabularnewline
\hline 
\hline 
\textcolor{red}{00}0 & \textcolor{red}{00}1\tabularnewline
\hline 
\textcolor{red}{01}1 & \textcolor{red}{01}0\tabularnewline
\hline 
\textcolor{red}{11}0 & \textcolor{red}{11}1\tabularnewline
\hline 
\textcolor{red}{10}1 & \textcolor{red}{10}0\tabularnewline
\hline 
\end{tabular}\caption{ Binary representations of $G_{3}$ divided into even and odd terms.}\label{tab2}
\end{table}

According to Eq. \ref{eq1}, the operator $b^{2}$ in this encoding
($G_{n}$) can be expressed as:

\begin{align}
b^{2} & =\sum_{i=1}^{N-1}\sqrt{i}\mathcal{X}_{i,i-1}'X_{0}\mathcal{P}_{i}X_{0}\sum_{j=1}^{N-1}\sqrt{j}\mathcal{X}_{j,j-1}'\mathcal{P}_{j}\nonumber \\
 & =\left(\sum_{i=1}^{N-1}(-1)^{i_{0}}\sqrt{i}\mathcal{X}_{i,i-1}'\mathcal{P}_{i}\right)\left(\sum_{j=1}^{N-1}\sqrt{j}\mathcal{X}_{j,j-1}'\mathcal{P}_{j}\right)\label{eq7}
\end{align}
where $\mathcal{X}_{i,i-1}'$ represents the collection excluding
the $0$-th qubit position, and $i_{0}$denotes the value of the 0-th
bit in the binary representation of $i$. The second line in Eq. \ref{eq7}
is derived by utilizing the identity $X_{0}Z_{0}^{\delta}X_{0}=(-1)^{\delta}Z_{0}^{\delta}$
for any $\delta$. This expression demonstrates that if there is an
$X$ operator acting on each term of $\mathcal{X}_{i,i-1}$ at a given
bit position, the number of terms in the operator's expression remains
unchanged. In Appendix \ref{app2}, we shown that $G_{n}$ can be
used as any $2^{\xi}$-fold code, therefor $G_{n}$ is suit for encoding
the operator $b^{l}$ with $l=2^{\xi}$. 

\begin{figure}
\includegraphics[scale=0.55]{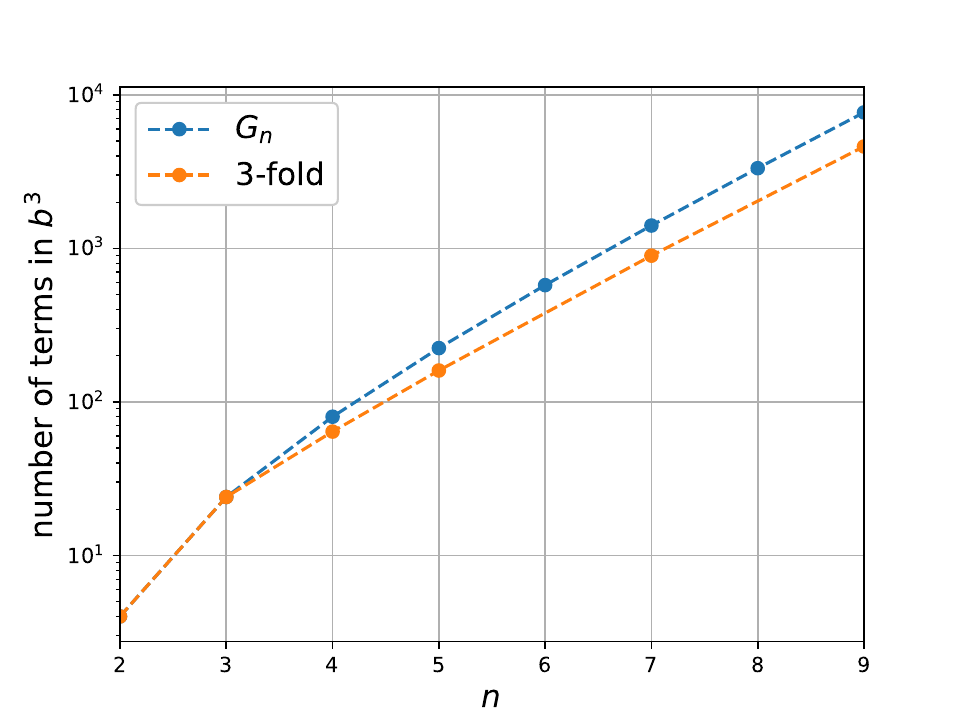}\caption{ Number of terms in the encoding operator $b^{3}$ as a function
of the number of qubits $n$. The cyna dashed line represents the
Gray code $G_{n}$, and the orange dashed line represents the 3-fold
code.}\label{fig2}
\end{figure}

Next, we present a counterexample to examine the limitations of $G_{n}$.
Specifically, when $k=3$, the local symmetry of $G_{n}$ is lost,
and a 3-fold code should be the most suitable for encoding in this
scenario. As shown in Fig. \ref{fig2}, we display the number of terms
in $b^{3}$ as a function of the number of qubits $n$. It is evident
that for $n\ge3$, $G_{n}$ is no longer the optimal encoding method.

It is important to note that the figure lacks data for the 3-fold
code at $n=6$ and $n=8$. This omission is due to the absence of
corresponding 3-fold codes for these values of $n$. Efficiently finding
3-fold codes is mathematically challenging, and thus, we do not discuss
this in further detail here.

\section{Squeezed state simulation}\label{secSSS}

\subsection{Encoded Hamiltonian}\label{subsec:EH}

For practical quantum simulation, we consider encoding the squeezed
state dynamics using a Gray code method. In previous studies \cite{squeeze1},
a one-hot encoding scheme was used to represent a single-mode squeezed
state. In a one-hot scheme with $n$ qubits, each basis state features
exactly one qubit in the $\ket 1$ state and the remainder in the
$\ket 0$ state, allowing the circuit to represent photonic states
with up to $n-1$ photons (yielding $n$ distinct states). In contrast,
Gray code encoding exploits the full $2^{n}$ computational basis,
enabling simulation of photonic states with up to $2^{n}-1$ photons.

The single-mode squeezed state can be generated by applying a squeezing
operator $S(z)$ to the vacuum state. The squeezing operator is defined
as
\begin{align}
S(z) & =\exp\left[\frac{1}{2}\left(z^{*}b^{2}-zb^{\dagger2}\right)\right],\label{eq10-1}
\end{align}
where $z=re^{i\varphi_{z}}$ is the complex squeezing parameter (with
$r=|z|$ characterizing the squeezing level and complex angle $\varphi_{z}$
determine the squeezing direction in the phase space). Here $b$ is
the photon annihilation operator for the mode. The squeezing process
can be viewed as an evolution under a Hamiltonian $H$: specifically
$S=e^{-iHt}|_{t=r},$ evaluated at time $t=r$. The Hamiltonian that
generates this transformation is given by
\begin{align}
H & =\frac{1}{2}\left[e^{-i\left(\varphi_{z}-\frac{\pi}{2}\right)}b^{2}-e^{i\left(\varphi_{z}+\frac{\pi}{2}\right)}b^{\dagger2}\right].\label{eq11}
\end{align}
Applying the operator $S(z)$ to the vacuum $|0\rangle$ (the single-mode
Fock vacuum) yields an analytical expression for the squeezed state
\cite{Vogel}, and it is given by 
\begin{align}
\ket z & =\frac{1}{\sqrt{\mu}}\exp\left(-\frac{\nu}{2\mu}b^{\dagger2}\right)\ket 0,\label{eq12}
\end{align}
where $\mu=\cosh r$ and $\nu=e^{i\varphi_{z}}\sinh r$. Notably,
the Fock basis expansion of $|z\rangle$ contains only even-photon-number
components. Exploiting this property, we improve our encoding efficiency
by representing only even-photon Fock states. Consequently, the $2^{n}$
available codewords can encode Fock states with photon numbers up
to $2(2^{n}-1)$; in other words, with $n$ qubits, we can represent
photon numbers as high as twice the usual limit.

For example, with $n=2$ qubits, a one-hot encoding can represent
up to $1$ photon, while a Gray code can normally represent up to
$3$ photons. If we restrict to even photon numbers, those two qubits
can instead represent the four Fock states $\{|0\rangle_{F},|2\rangle_{F},|4\rangle_{F},|6\rangle_{F}\}$
(0, 2, 4, and 6 photons). We assign these Fock states to the 2-qubit
computational basis states using the Gray code sequence $G_{2}=\{|00\rangle,|01\rangle,|11\rangle,|10\rangle\}$.
According to the general encoding formula Eq. \ref{eq7}, the annihilation
operator $b^{2}$ under this encoding is expressed as:
\begin{align}
b^{2} & =\frac{\sqrt{2}+\sqrt{30}}{4}X_{0}+\frac{\sqrt{12}}{4}X_{1}+\frac{\sqrt{2}-\sqrt{30}}{4}iY_{0}+\frac{\sqrt{12}}{4}iY_{1}\nonumber \\
 & +\frac{\sqrt{2}-\sqrt{30}}{4}Z_{1}X_{0}+\frac{\sqrt{2}+\sqrt{30}}{4}iZ_{1}Y_{0}\nonumber \\
 & -\frac{\sqrt{12}}{4}X_{1}Z_{0}-\frac{\sqrt{12}}{4}iY_{1}Z_{0}.
\end{align}
Substituting this encoded form of $b^{2}$ into the Hamiltonian $H$
in Eq. \ref{eq11}, we obtain the qubit-encoded Hamiltonian for the
squeezed state, and it is given by: $H=H_{R}+H_{I},$ where
\begin{align}
H_{R} & =\cos\varphi_{z}\left(\frac{\sqrt{30}-\sqrt{2}}{4}Y_{0}-\frac{\sqrt{3}}{2}Y_{1}\right.\label{eq14}\\
 & \left.-\frac{\sqrt{30}+\sqrt{2}}{4}Z_{1}Y_{0}+\frac{\sqrt{3}}{2}Y_{1}Z_{0}\right)\nonumber 
\end{align}
and 
\begin{align}
H_{I} & =\sin\varphi_{z}\left(\frac{\sqrt{30}+\sqrt{2}}{4}X_{0}+\frac{\sqrt{3}}{2}X_{1}\right.\label{eq15-1}\\
 & \left.-\frac{\sqrt{30}-\sqrt{2}}{4}Z_{1}X_{0}-\frac{\sqrt{3}}{2}X_{1}Z_{0}\right).\nonumber 
\end{align}
 Thus, we have explicitly encoded the squeezing Hamiltonian in terms
of two-qubit operations. This encoded Hamiltonian can now be used
in the circuit model to simulate the squeezed state dynamics on a
quantum processor.

\subsection{Variational quantum simulation}\label{subsec:VQS}

For near-term quantum devices where noise and limited gate depth hinder
deep circuit implementations, the traditional Suzuki--Trotter method
becomes impractical. Instead, we adopt Variational Quantum Simulation
(VQS) \cite{VQS1,VQS2,VQS3}, which uses a shallow, parameterized
circuit to approximate time evolution. We provide here a concise introduction
to the principle of VQS.

\subsubsection{Ansatz for evolution}

Once the physical system is encoded into a circuit model, the Hamiltonian
can be expressed as a weighted sum of Pauli strings,
\begin{align}
H & =\sum_{i}\xi_{i}P_{i},
\end{align}
where each $\xi_{i}$ is a real coefficient and each operator $P_{i}$
is given by a tensor product of Pauli matrices (i.e. $P_{i}=\sigma_{n-1}\otimes...\otimes\sigma_{0}$
with $\sigma_{i}\in\{I,X,Y,Z\}$ acting on the corresponding qubit
in the circuit). Given an initial state $\ket{\psi_{0}}$, the exact
time-evolved state under $H$ is
\begin{align*}
\ket{\psi_{t}} & =e^{-iHt}\ket{\psi_{0}}.
\end{align*}
In the fault-tolerant era, one standard approach is the Suzuki--Trotter
decomposition:
\begin{align*}
e^{-iHt} & \approx\left(\prod_{j}e^{-iP_{j}\xi_{j}\frac{t}{N_{\tau}}}\right)^{N_{\tau}},
\end{align*}
with large $N_{\tau}$. However, on near-term quantum devices subject
to noise and gate imperfections, large circuit depths quickly degrade
the fidelity of the final state. A variational strategy addresses
this by approximating the time evolution via a shallower, parametrized
circuit. This method is often referred to as VQS \cite{VQS1,VQS2,VQS3}---an
approach related to the philosophy of variational quantum eigensolvers
(VQE) \cite{VQE_first,VQE_first_review}. 

In a VQS approach, one replaces the exact propagator $e^{-iHt}$ by
a parametrized, shallower-depth ansatz circuit
\begin{align}
\Gamma(\boldsymbol{\theta}) & =\Gamma(\theta_{0},\theta_{1},...,\theta_{N-1}),
\end{align}
where each $\theta_{i}=\theta_{i}(t)$ is a time-dependent variational
parameter. The total number of these parameters is kept deliberately
small to maintain low-depth circuits. The goal is then to evolve the
system in time by adjusting $\boldsymbol{\theta}(t)$ so that
\begin{align}
\ket{\psi_{t}} & \approx\ket{\psi(\boldsymbol{\theta})}=\Gamma(\boldsymbol{\theta})\ket{\psi_{0}}.\label{eq25}
\end{align}

\subsubsection{Equations of motion (EOM) of VQS}

This strategy aims to capture the essential dynamics of the system
without incurring the large circuit depths required by direct Trotter
decomposition. The time-dependent variational principle which derives
the Schrödinger equation \cite{VQS1} is given by 
\begin{align}
\delta\int dtL & =0,\label{eq26}
\end{align}
where the $L=\bra{\psi(\boldsymbol{\theta})}i\frac{\partial}{\partial t}-H\ket{\psi(\boldsymbol{\theta})}$
is the Lagrangian of the system. With considering Eq. \ref{eq25},
the Lagrangian in this representation is given by 
\begin{align}
L(\boldsymbol{\theta},\dot{\boldsymbol{\theta}}) & =i\sum_{k}\bra{\psi(\boldsymbol{\theta})}\frac{\partial\ket{\psi(\boldsymbol{\theta})}}{\partial\theta_{k}}\dot{\theta}_{k}-\bra{\psi(\boldsymbol{\theta})}H\ket{\psi(\boldsymbol{\theta})}.
\end{align}
Therefore the variational principle in Eq. \ref{eq26} derives the
equation set 
\begin{align}
\sum_{q}M_{k,q}\dot{\theta}_{q} & =V_{k},\label{eqMV}
\end{align}
where 
\begin{align}
M_{kq} & =i\eta\frac{\partial\bra{\psi}}{\partial\theta_{k}}\frac{\partial\ket{\psi}}{\partial\theta_{q}}+h.c.\label{eqM}
\end{align}
\begin{align}
V_{k} & =\eta\frac{\partial\bra{\psi}}{\partial\theta_{k}}H\ket{\psi}+h.c.\label{eqV}
\end{align}
with $\eta=1$. Here, “h.c.” stands for Hermitian conjugate. These
equations are not numerical stable \cite{VQS3}. The matrix $\boldsymbol{M}$
( whose elements are $M_{kq}$) is antisymmetric, leading to $\det\boldsymbol{M}=0$
whenever the number of variational parameters is odd---hence no unique
solution. And there is no practical one-parameter ansatz circuit in
that simple form, which is very useful in many cases such as debugging.
A more stable way is to use the McLachlan’s variational principle
\cite{BROECKHOVE1988547}. It derives the same equation set in Eq.
\ref{eqM} and Eq. \ref{eqV} except that $\eta=-i$. In this case,
the matrix $\boldsymbol{M}$ is symmetric. We adopt this formulation
in our subsequent calculations. 

\subsubsection{Circuit for calculation $M$ and $V$}

\begin{figure}
\begin{quantikz}
\lstick{$|0\rangle$}&\gate{H}&\octrl{1}& &\ctrl{1}&\gate{H}&\meter{}\\
\lstick{$|\psi_0\rangle$}&\gate{\Gamma_{0}^{k-1}}&\gate{\sigma_k}&\gate{\Gamma_{k}^{q-1}}&\gate{\sigma_q}& & \\
\end{quantikz}
\caption{ The ancilla‐based measurement circuit used to estimate each matrix
element $M_{kq}$. }\label{figM}
\end{figure}
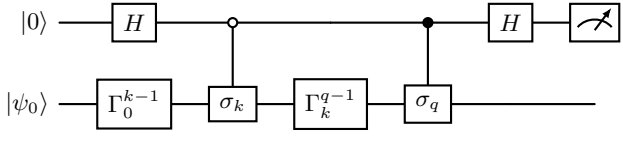

We consider a layered or factorized ansatz of the form
\begin{align*}
\Gamma(\theta) & =\Gamma_{N-1}(\theta_{N-1})\Gamma_{N-2}(\theta_{N-2})...\Gamma_{0}(\theta_{0}),
\end{align*}
where each $\Gamma_{i}(\theta_{i})=A_{i}R_{i}(\theta_{i})$ is composed
of:
\begin{enumerate}
\item A fixed (time‐independent) product of gates $A_{i}$
\item A single‐qubit rotation $R_{i}(\theta_{i})=e^{-i\sigma_{i}\frac{\theta_{i}}{2}}$
about a specified axis $\sigma_{i}\in\{X,Y,Z\}$.
\end{enumerate}
It follows that 
\begin{align}
\frac{d}{d\theta_{i}}R_{i}(\theta_{i}) & =\frac{1}{2i}R_{i}\sigma_{i}.
\end{align}
From this, one can derive
\begin{align}
\frac{\partial\Gamma(\boldsymbol{\theta})}{\partial\theta_{i}} & =\Gamma_{N-1}...\Gamma_{i+1}\Gamma_{i}\sigma_{i}\Gamma_{i-1}...\Gamma_{0}.\label{eqGamma}
\end{align}
Using the above derivative rule in the definition of $M_{kq}$ (Eq.
\ref{eqM}), we obtain

\begin{align}
M_{kq} & =\frac{i}{4}\bra{\psi_{0}}\Gamma_{0}^{k-1\dag}\sigma_{k}^{\dagger}\Gamma_{k}^{q-1\dagger}\sigma_{q}\Gamma_{k}^{q-1}\Gamma_{0}^{k-1}\ket{\psi_{0}}+h.c.,\label{eqM2}
\end{align}
where we have used the notation $\Gamma_{k}^{k+l}=\Gamma_{k+l}...\Gamma_{k+1}\Gamma_{k}$.
To measure $M_{kq}$ on a quantum processor, one can implement the
circuit in Fig.\, \ref{figM}, where an ancillary qubit is introduced
and used to perform a measurement‐based estimation of the above amplitude.
Denoting the measured probabilities of the ancilla as $p_{0}$ and
$p_{1}$ (for observing $0$ and $1$ respectively), one has
\begin{align}
M_{kq} & =\frac{1}{2}(p_{0}-p_{1})=p_{0}-\frac{1}{2}.
\end{align}
This procedure is repeated for all required pairs $(k,q)$ to construct
the full matrix $\boldsymbol{M}$. 

\begin{figure}
\begin{quantikz}
\lstick{$|0\rangle$}&\gate{H}&\octrl{1}& &\ctrl{1}&\gate{H}&\meter{}\\
\lstick{$|\psi_0\rangle$}&\gate{\Gamma_{0}^{k-1}}&\gate{\sigma_k}&\gate{\Gamma_{k}^{n-1}}&\gate{P_m}& &\\
\end{quantikz}
\caption{ Circuit diagram for measuring the vector components $V_{km}$. }\label{figV}
\end{figure}
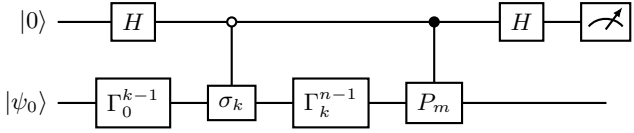

Similarly, one obtains from the definition of $V_{k}$ (Eq. \ref{eqV})
that $V_{k}=2\sum_{m}\xi_{m}V_{km},$ where

\begin{align}
V_{km} & =\frac{i}{4}\bra{\psi_{0}}\Gamma_{0}^{k-1\dagger}\sigma_{k}\Gamma_{k}^{n-1\dagger}P_{m}\Gamma_{k}^{n-1}\Gamma_{0}^{k-1}\ket{\psi_{0}}+h.c..\label{eqV2}
\end{align}
As with $M_{kq}$, the circuit for estimating $V_{km}$ is nearly
the same, except that $P_{m}$ is inserted in place of $\sigma_{q}$.
This is illustrated in Fig.\,\ref{figV}. 

\subsubsection{Variational Hamiltonian ansatz (VHA)}

In the preceding discussions, we introduced the VQS framework and
highlighted that one must specify the functional form of the variational
ansatz $\Gamma(\boldsymbol{\theta})$. Choosing this ansatz is a nontrivial
challenge, as naive or random circuit structures can lead to two major
problems. First, a poorly chosen ansatz may yield trivial parameter
dynamics, producing $\dot{\boldsymbol{\theta}}(t)=0$ for all $t$
and implying no meaningful evolution. Second, even when the EOM do
evolve in time, a suboptimal ansatz might fail to capture key features
of the quantum state, so that no matter how small the time‐step $\Delta t$
we used in incrementing parameters, the resulting simulation deviates
substantially from the true $e^{-iHt}\ket{\psi_{0}}$. These issues
have been extensively discussed in Ref.\, \cite{VQS3}. One robust
strategy to mitigate them is the so‐called Variational Hamiltonian
Ansatz (VHA) \cite{VHA1,VHA2}. In VHA, we construct the parametrized
circuit by mimicking short Trotter‐like steps of the original Hamiltonian.
Concretely, we consider a circuit with multiple layers (“depth” $n_{d}$)
of exponentials of each Pauli string. That is, we write:
\begin{align}
\Gamma(\boldsymbol{\theta}) & =\prod_{d=0}^{n_{d}-1}\left[\prod_{j}\exp\left(-iP_{j}\theta_{dj}\right)\right].
\end{align}
Each layer thus contains exponentials of the same set of Pauli operators
$\{P_{j}\}$ appearing in $H$. In effect, one can view each layer
as a “Trotter slice,” except that we allow different, variationally
optimized angles $\theta_{dj}$ per layer. This approach has several
advantages. It is systematically improvable: by increasing $n_{d}$,
one enlarges the variational space and can asymptotically recover
exact dynamics. It also has physical interpretability: each parameter
$\theta_{dj}$ directly corresponds to how strongly the relevant Hamiltonian
term acts. Moreover, it reduces the risk of stagnation because the
VQS solution coincides with the exact propagator for vanishing small
times \cite{VQS3}. 

In practice, one sets the desired number of layers $n_{d}$ by starting
with the smallest possible $(n_{d}=1)$ and incrementing until the
approximation quality is acceptable. We adopt the VHA in our calculations. 

\subsubsection{Summary of the VQS procedure}

The steps to implement VQS are concluded as follows:
\begin{enumerate}
\item Initialization: Set $\boldsymbol{\theta}(t=0)$ so that $\Gamma(\boldsymbol{\theta})=\boldsymbol{1}$
is effectively the identity. Therefore we have $\ket{\psi_{t=0}}=\ket{\psi_{0}}$.
\item Measurement of $\boldsymbol{M}$ and $\boldsymbol{V}$: For the current
parameter set $\boldsymbol{\theta}(t)$, run the ancilla-aided circuits
in Fig. \ref{figM} and Fig. \ref{figV} to obtain all elements $M_{kq}$
and $V_{k}$. Build the matrix $\boldsymbol{M}$ and vector $\boldsymbol{V}$
from these measurements.
\item Update rule: Solve the linear system $\boldsymbol{M}(t)\dot{\boldsymbol{\theta}}(t)=\boldsymbol{V}(t)$
for $\dot{\boldsymbol{\theta}}(t)$. In practice, one can use standard
linear algebra methods. 
\item Increment parameters: Propagate by a small time step $\Delta t$,
i.e. $\boldsymbol{\theta}(t+\Delta t)\approx\boldsymbol{\theta}(t)+\dot{\boldsymbol{\theta}}(t)\Delta t$.
After this, go to step 2 to start new calculation for the next time
step.
\item Approximate the evolved state: At any intermediate time $t$, the
circuit $\Gamma(\boldsymbol{\theta})$ approximates the true evolution
$e^{-iHt}$. Thus $\ket{\psi_{t}}\approx\Gamma\left(\boldsymbol{\theta}(t)\right)\ket{\psi_{0}}$. 
\end{enumerate}

\section{Results}\label{sec:results}

\subsection{Fidelity of VQS}

\begin{figure}
\includegraphics[scale=0.55]{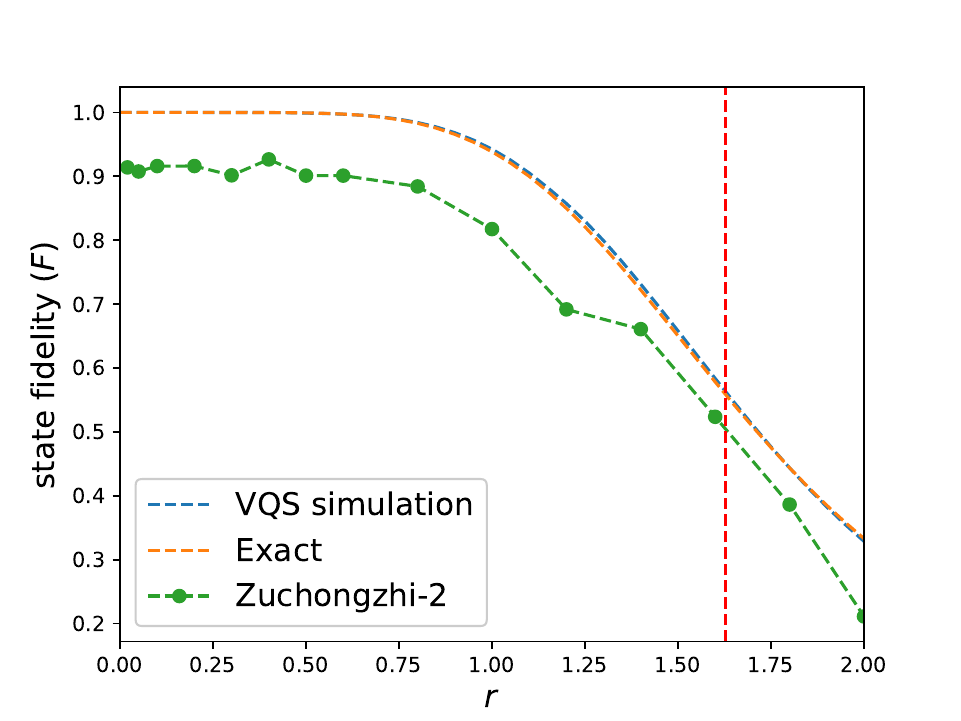}\caption{ Fidelity vs. $r=|z|$ for the truncated exact solution (blue dashed),
VQS simulation (blue dashed), and quantum hardware \textit{Zuchongzhi-2}
(green). The vertical red dashed line marks $r_{0}\approx1.63$, where
$n^{s}(r_{0})=6$.}\label{fig3}
\end{figure}

We employ the Gray code encoding to simulate the evolution of a vacuum
state under single-mode squeezing. As discussed before, the encoding
is restricted to even-photon Fock states, thus omitting contributions
from odd photon numbers. In the illustrative examples presented here,
the truncated subspace spans four Fock states $\{\ket 0_{F},\ket 2_{F},\ket 4_{F},\ket 6_{F}\}$,
allowing us to capture photon numbers up to six. For simplification,
the squeezing parameter $z$ is taken to be purely imaginary ($z=it$),
so that the real component of the squeezing Hamiltonian vanishes.
Empirically, a single variational layer ($n_{d}=1$) suffices for
accurately capturing the relevant dynamics.

To assess simulation accuracy, we define the fidelity of a state $\ket{\psi}$
(restricted to this four dimensional subspace) as $F(\ket{\psi})=|\braket{\psi}z|^{2}$,
where $\ket z$ is the exact squeezed state (see Eq. \ref{eq12}),
truncated to the same maximum photon number of six in this case. 

Figure \ref{fig3} plots the fidelity versus $r=|z|$. The curve labeled
“Exact” represents the mathematical wavefunction of the squeezed state,
forcibly truncated at six photons. Because a true single-mode squeezed
state has an average photon number $n^{s}(z)=\sinh^{2}r$, once $n^{s}(z)$
grows beyond $6$, the truncated subspace no longer captures a significant
portion of the total wavefunction. Consequently, this truncated “Exact”
fidelity drops below unity for large $r$. A vertical red dashed line
indicates $r_{0}\approx1.63$, where $n^{s}(r_{0})=6$; beyond that
point, truncating at six photons omits a substantial fraction of the
state.

The VQS (blue dashed line) closely follows the truncated exact curve
across nearly the entire range. Only near $r\approx1.25$ does a minor
discrepancy arise. Even then, the deviation remains small, underscoring
that our Gray code encoding and single-layer variational approach
can reliably capture squeezed-state evolution up to quite large $r$.
For $r$ beyond the red dashed line, the average photon number exceeds
six, causing the truncated state (and hence our benchmark) to lose
fidelity more substantially.

\subsection{Quantum state tomography}

\begin{figure}
\includegraphics[scale=0.55]{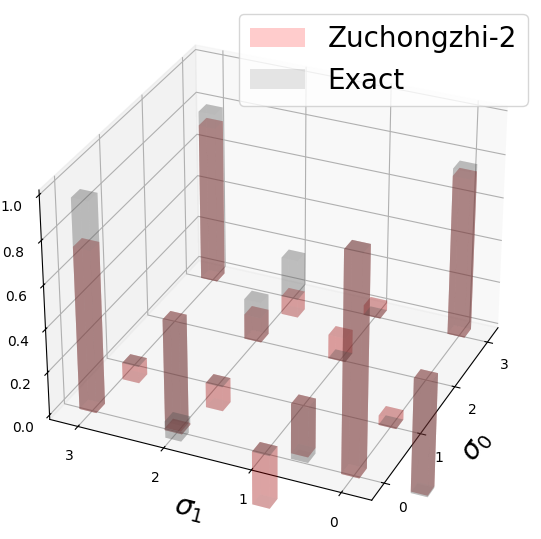}\caption{ Reconstructed two\protect\nobreakdash-qubit density matrix $\rho$
at $r=0.5$. Red bars represent the measured data (via quantum state
tomography on the \textit{Zuchongzhi\protect\nobreakdash-2} processor)
through the cloud platform, whereas gray bars show the corresponding
truncated theory values.}\label{fig4}
\end{figure}

To quantify how accurately our quantum processor reproduces the intended
squeezed state, we perform quantum state tomography (QST) on the final
two\nobreakdash-qubit state. For a two\nobreakdash-qubit system
in SU(2)\ensuremath{\otimes}SU(2), we can decompose the density matrix
$\rho$ into a sum of tensor\nobreakdash-product Pauli operators
\cite{Nielsen2010}:
\begin{align}
\rho & =\sum_{ij}\frac{\tr{\left(\rho\sigma_{i}^{1}\otimes\sigma_{j}^{0}\right)}}{4}\sigma_{i}^{1}\otimes\sigma_{j}^{0},\label{eq21-1}
\end{align}
where each coefficient $\tr{\left(\rho\sigma_{i}^{1}\otimes\sigma_{j}^{0}\right)}$
is determined via measurements in the corresponding Pauli basis. Although
the total number of measurement settings scales exponentially with
qubit count, tomography remains tractable for two qubits. For larger
systems, an ancilla\nobreakdash-based Hadamard test \cite{quantumalgorithms}
may be more efficient, but it requires extra qubits and additional
entangling gates, which can introduce further noise.

In our demonstration, we implement direct measurements of each two\nobreakdash-qubit
Pauli basis on the \textit{Zuchongzhi-2} device, collecting 50,000
shots per basis to bolster the statistical reliability (see Appendix
\ref{hardware} for more details of quantum hardware). We further
optimize performance by transpiling the parameterized circuit (with
the calibrated parameters $\boldsymbol{\theta}$) into \textit{Zuchongzhi-2}’s
native gate set $\{X2P,X2M,Y2P,Y2M,R_{z},CZ\}$ using \textit{bqskit}
\cite{BQSKit}, thus minimizing gate layers and reducing error accumulation.
In Appendix \ref{Transpile}, we will give an example of this calculation.

Figure \ref{fig4} displays the reconstructed two\nobreakdash-qubit
density matrix $\rho$ for $r=0.5$. Although certain components,
such as $Z\otimes I$ and $Y\otimes Z$, deviate noticeably from their
ideal values (e.g., in amplitude and sign), the overall fidelity $\tr{\left(\rho\ket z\bra z\right)}$
remains above $0.9$. This indicates that the core features of the
squeezed state are captured relatively well. Referring back to Fig.
\ref{fig3} (green dot\nobreakdash-dashed line), we see that across
a broad range of $r$, the hardware fidelity remains high, agreeing
well with the ideal truncated solution until the effective photon
number grows too large for our 4\nobreakdash-dimensional qubit encoding.
These results confirm that, even with hardware imperfections and finite
sampling, the proposed encoding and variational approach faithfully
reproduce a significant portion of the squeezed state’s target distribution.

\subsection{Wigner spectra }

\begin{figure*}
\includegraphics[scale=0.5]{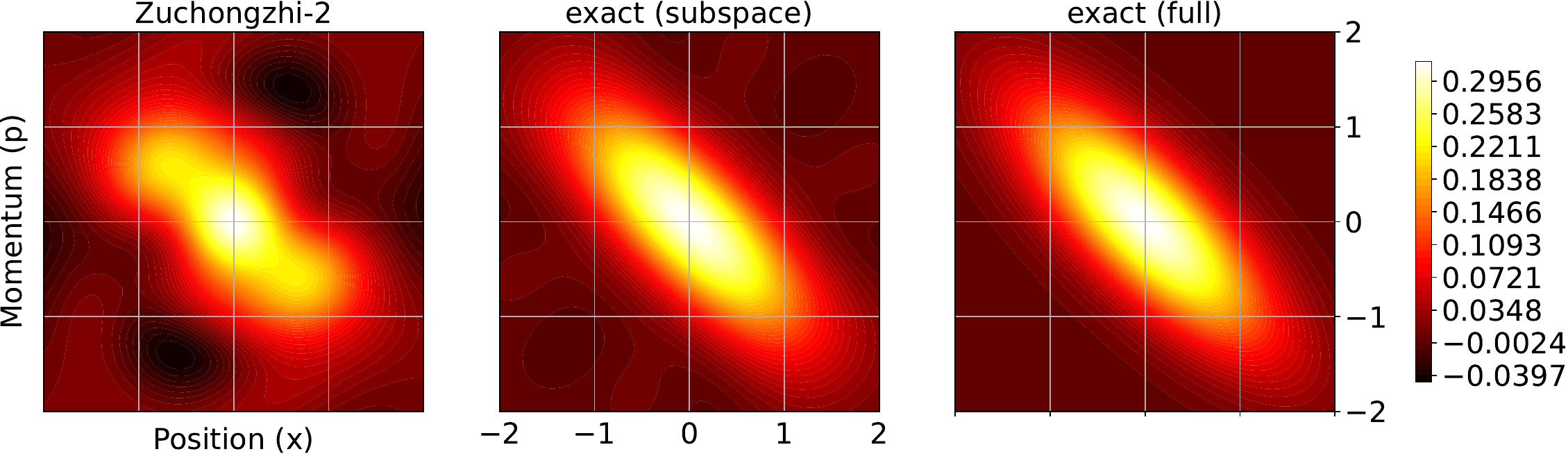}\caption{ Wigner distributions of the single-mode squeezed state at $|z|=0.5$.
From left to right: (i) Hardware demonstration result on \textit{Zuchongzhi-2}
superconducting quantum processor, (ii) theoretical (exact) truncation
to a six-photon Fock subspace, and (iii) the ideal infinite-dimensional
state.}\label{fig5}
\end{figure*}

Once the two-qubit density matrix has been reconstructed via QST,
we can evaluate the corresponding Wigner-function \cite{wigner,qutip}
to visualize the system’s phase-space characteristics. Figure \ref{fig5}
illustrates the Wigner distribution at $r=0.5$ for three cases (from
the left to the right in the figure): (i) the hardware demonstration
result on the QPU \textit{Zuchongzhi-2}, (ii) the theoretical (exact)
result truncated to a six-photon Fock subspace, and (iii) the exact
infinite-dimensional squeezed state.

We can factor the single-mode squeezing operator as $S(z)=R(\frac{\varphi_{z}}{2})S(r)R(-\frac{\varphi_{z}}{2})$,
where $S(r)$ compresses the $x$-quadrature (and stretches $p$).
Since $\varphi_{z}=\frac{\pi}{2}$ here, the overall operation is
rotated by $\frac{\pi}{4}$, leading to the 45° tilt in the Wigner
plots, as shown in Fig. \ref{fig5}. 

From the infinite-dimensional perspective, a squeezed state at this
moderate squeezing value exhibits a purely positive Wigner-function.
The finite truncation at six photons, however, introduces slight negative
“ring-like” regions toward the fringes, reflecting the imperfect capture
of higher photon-number components. These negative pockets do not
represent genuine nonclassical phenomena at this moderate squeezing
level; rather, they are an artifact of disallowing photon numbers
above six.

Through cloud platform demonstration, our reconstructed Wigner-function
largely retains the squeezed profile but shows more widespread negative
dips than the truncated case. This discrepancy underscores the practical
effects of both device noise and the same finite-photon cutoff. Nevertheless,
the central squeezed peak and its diagonal orientation are readily
discernible, indicating that the main features of the targeted squeezed
state remain intact under our encoding and simulation procedure.

Overall, these Wigner spectra confirm that, despite hardware imperfections
and photon-number truncation, the essential squeezed-state characteristics
(notably, the elongated shape in phase space) are successfully reproduced.
This consistency with theoretical expectations, even for a relatively
modest number of computational basis states, further validates our
bosonic encoding strategy and demonstrates the feasibility of exploring
continuous-variable phenomena on digital quantum processors.

\section{Conclusion}\label{sec:conclusion}

In this work, we have developed a comprehensive framework to encode
$2^{n}$ bosonic modes using $n$ qubits. Within this paradigm, the
Gray code stands out by achieving excellent efficiency---minimizing
the number of Pauli strings needed to represent the operator $b^{l}$
(with $l=2^{\xi}$)---due to its inherent symmetry. This efficiency
directly translates into reduced circuit depth and lower gate complexity.

Furthermore, we introduced a truncated encoding scheme that maps only
even-photon Fock states, thereby effectively doubling the maximum
photon number accessible per qubit. For instance, with $n=2$ qubits,
our method can simulate photonic states with up to $6$ photons. This
encoding strategy was integrated into a variational quantum simulation
(VQS) framework, where shallow circuits and ancilla-assisted measurements
were used to iteratively update the time-dependent variational parameters
($M_{kq}$ and $V_{km}$). The VQS results show good agreement with
exact simulations, demonstrating the robustness of this approach.

Demonstration implementations on the Zuchongzhi-2 processor further
validated our method by producing high-fidelity squeezed states at
moderate squeezing levels, as confirmed by both quantum state tomography
and Wigner-function measurements. Overall, our results not only establish
the feasibility of simulating continuous-variable (CV) states on digital,
qubit-based quantum processors but also underscore the promising potential
of Gray-code methods for a wide range of bosonic Hamiltonians.
\begin{acknowledgments}
This work was supported by the Beijing Nova Program (Grants No. 20220484128
and 20240484652). Computational resources were provided by the Chaohumingyue
Hefei QC-HPC Hybrid Computing Center. All of our numerical calculations
are based on \textit{isQ} \cite{10129229}.
\end{acknowledgments}

\appendix

\section{Projector }\label{appendix:Projector}

We begin with the expression for the annihilation operator in the
Fock basis:
\begin{align}
b & =\sum_{i=1}^{N-1}\sqrt{i}\ket{i-1}_{F}\bra i_{F}.
\end{align}
Consider the encoding mapping $\{\ket{c_{i}}\}$. Under this encoding,
the operator $\ket{i-1}_{F}\bra i_{F}$ is mapped to:
\begin{align}
\ket{i-1}_{F}\bra i_{F} & =\ket{c_{i-1}}\bra{c_{i}}\nonumber \\
 & =O_{n-1}\otimes O_{n-2}\otimes...\otimes O_{0},\label{eq9}
\end{align}
where $O_{\alpha}$ is an operator acting on the $\alpha$-th qubit.
The explicit form of $O_{\alpha}$ depends on the values of $c_{i-1}$
and $c_{i}$ at the $\alpha$-th bit of their binary representations.
Specifically, $O_{\alpha}$ is defined as \cite{Mohan_2025}:
\begin{equation}
O_{\alpha}=\left\{ \begin{array}{c}
\ket 0\bra 0=(\boldsymbol{1}+Z)/2\text{ if }c_{i-1}^{\alpha}=0\text{ and }c_{i}^{\alpha}=0\\
\ket 1\bra 1=(\boldsymbol{1}-Z)/2\text{ if }c_{i-1}^{\alpha}=1\text{ and }c_{i}^{\alpha}=1\\
\ket 1\bra 0=(X-iY)/2\text{ if }c_{i-1}^{\alpha}=1\text{ and }c_{i}^{\alpha}=0\\
\ket 0\bra 1=(X+iY)/2\text{ if }c_{i-1}^{\alpha}=0\text{ and }c_{i}^{\alpha}=1
\end{array}\right..\label{eq8}
\end{equation}

We can simplify the last two cases to obtain a more compact form.
For example, the operator $\ket 1\bra 0$ can be rewritten as:
\begin{align}
\ket 1\bra 0 & =\frac{1}{2}(X-ZX)\nonumber \\
 & =\frac{1}{2}X(\boldsymbol{1}+Z)=X\cdot\ket 0\bra 0.
\end{align}
This demonstrates that the transformation operator $\ket 1\bra 0$
can be expressed as a projection onto the $\ket 0$ state followed
by a bit flip $X$ on the relevant qubit. Similarly, we have: $\ket 0\bra 1=X\cdot\ket 1\bra 1.$
Therefore $\ket{c_{i-1}}\bra{c_{i}}$ can be written as: 
\begin{align}
\ket{c_{i-1}}\bra{c_{i}} & =\mathcal{X}_{i-1,i}\mathcal{P}_{i},
\end{align}
where $\mathcal{P}_{i}=\ket{c_{i}}\bra{c_{i}}$ is the projector onto
the state $\ket{c_{i}},$ and $\mathcal{X}_{i-1,i}$ is defined as
described in the main text. The state projector $\mathcal{P}_{i}$
can be further expressed in terms of Pauli $Z$ operators as: 
\begin{align}
\mathcal{P}_{i} & =\frac{1}{2^{n}}\left[\boldsymbol{1}+(-1)^{i_{n-1}}Z\right]\otimes...\otimes\left[\boldsymbol{1}+(-1)^{i_{0}}Z\right],
\end{align}
where $i_{\alpha}$ is the value of the $\alpha$-th bit (0-based)
of the integer $i$ in its binary representation. Expanding the tensor
products yields Eq. \ref{eq3} from the main text.

\section{$2^{\xi}$-fold code in Graycode }\label{app2}

The Gray code $G_{n}$ is divided into $l=2^{\xi}$ groups, where
$\xi\le n-1$. Each group consists of elements from $G_{n}$ that
are spaced $l$ apart. Specifically, for each $l$ , the Gray code
$G_{n}$ is partitioned such that the $i$-th group contains all elements
of $G_{n}$ with indices satisfying $j\equiv i\mod l$, for $i=0,1,...,l-1$.
Each group contains $2^{n-\xi}$ elements. We define the numbers within
a group to form a \textit{Gray code shape} if, aside from $\xi$ fixed
bit positions, the remaining bits constitute a Gray code. The values
of the fixed bits can either remain constant or undergo a single bit
flip (0 \ensuremath{\leftrightarrow} 1) between any two consecutive
numbers in the sequence. We will prove by induction that each such
group forms a Gray code shape. Consequently, this demonstrates that
$G_{n}$ is suitable for encoding the operator $b^{l}$. 

For $n=1$, the Gray code $G_{1}=(0,1).$ Dividing it into $l=2^{\xi}$
groups with $\xi=0,$ we get a single group containing $2^{1}=2$
elements. Clearly, each group trivially forms a Gray code shape. Assume
that for some $n=k$, the Gray code $G_{k}$ can be partitioned into
$l=2^{\xi}$groups, where each group forms a Gray code shape. Now,
consider $n=k+1.$ The Gray code $G_{k+1}$ is constructed using the
recursive formula $G_{k+1}=(\boldsymbol{0}\cdot G_{k},\boldsymbol{1}\cdot\bar{G}{}_{k})$.
By the inductive hypothesis, we know that $G_{k}$ is partitioned
into $l=2^{\xi}$ groups, and each group forms a Gray code shape.
When we extend to $G_{k+1}$, the same partitioning scheme applies.
Each of the $2^{\xi}$ groups in $G_{k}$ gives rise to two corresponding
groups in $G_{k+1}$ by by appending $0$ or $1$ to the front of
the elements in $G_{k}$. The bit flip between the two groups in $G_{k+1}$
only affects the leading bit, leaving the remaining bits from $G_{k}$
unaffected. Therefore, each of the $l$ groups is extended by directly
connecting the corresponding group in $\boldsymbol{0}\cdot G_{k}$
and $\boldsymbol{1}\cdot\bar{G}{}_{k}$. 
\begin{itemize}
\item Base Case ($n=1$)

For $n=1$, the Gray code is given by: $G_{1}=(0,1)$. Here, we must
have $\xi=0$ (since $\xi\le n-1$), so $l=2^{0}=1$. Thus, the entire
Gray code forms a single group containing $2^{1}=2$ elements. In
this case, the group trivially forms a Gray code shape since the two
elements differ in exactly one bit.
\item Inductive Hypothesis

Assume that for some $n=k$, the Gray code $G_{k}$ can be partitioned
into $l=2^{\xi}$ groups such that in each group, aside from $\xi$
fixed bits, the remaining $k-\xi$ bits form a Gray code. That is,
in every group, consecutive elements differ in exactly one bit in
the non-fixed positions, and any change in the fixed positions occurs
as a single bit flip.
\item Inductive Step ($n=k+1$)

The $(k+1)$-bit Gray code is constructed using the recursive formula:
$G_{k+1}=(\boldsymbol{0}\cdot G_{k},\boldsymbol{1}\cdot\bar{G}{}_{k})$.
We partition $G_{k+1}$ into $l=2^{\xi}$ groups by taking elements
with indices congruent modulo $l$. Notice that:
\begin{enumerate}
\item Within the First Half $(\boldsymbol{0}\cdot G_{k})$: Each group in
$G_{k}$(by the inductive hypothesis) forms a Gray code shape. Prefixing
these elements with $0$ merely adds a fixed bit to the beginning
of each element without affecting the Gray code property in the remaining
$k-\xi$ bits.
\item Within the Second Half $(\boldsymbol{1}\cdot\bar{G}{}_{k})$: Although
this half is constructed from the reversed $G_{k}$, the reversal
does not alter the property that consecutive elements differ in one
bit in the non-fixed positions. Prefixing these elements with 1 similarly
introduces a fixed bit that changes only in a controlled manner (at
most once between consecutive elements).
\item At the Junction Between the Two Halves: Consider the transition from
the last element of $\boldsymbol{0}\cdot G_{k}$ to the first element
of $\boldsymbol{1}\cdot\bar{G}{}_{k}$ within the same group. By the
standard construction of Gray codes, this transition involves a flip
in the leading bit only, while the remaining $k$ bits (inherited
from $G_{k}$ or its reverse) still satisfy the Gray code property
(i.e., they differ in exactly one bit as dictated by the inductive
hypothesis).
\end{enumerate}
\end{itemize}
Thus, in $G_{k+1}$, each of the $2^{\xi}$ groups is obtained by
“extending” the corresponding group from $G_{k}$ in both halves (with
the 0-prefix and 1-prefix, respectively). The only additional difference
between the two halves is in the newly added leading bit, and its
change is isolated to a single bit flip. Therefore, aside from the
$\xi$ fixed bits (now including the additional fixed prefix in each
half), the remaining bits in each group form a Gray code sequence.

\section{ Quantum Processor Specifications}\label{hardware}

The calculation of the demonstration were conducted on the \textit{Zuchongzhi-2}
superconducting quantum processor (66 qubits, topology shown in Fig.
8) via the \href{https://qc.zdxlz.com/home?lang=en}{Tianyan Quantum Computing platform}
(March 4--6, 2025). Qubits 21 ($f_{10}=5.1414$ GHz) and 26 ($f_{10}=5.2574$
GHz) were selected for their enhanced coherence properties, with measured
relaxation times $T_{1}=42.5285$ \textgreek{μ}s and $22.856$ \textgreek{μ}s,
respectively, and coherence times $T_{2}=2.5943$ \textgreek{μ}s and
$2.3345$ \textgreek{μ}s---compared to the chip-wide median $T_{1}=23.4$
\textgreek{μ}s and $T_{2}=3.38$ \textgreek{μ}s. Single-qubit gate
errors for these qubits measured $0.08$\% (qubit 21) and $0.12$\%
(qubit 26), outperforming the chip-average single-qubit error of $0.24$\%.
The CZ gate between qubits 21 and 26 (mediated by coupler G38) exhibited
a $1.04$\% error rate, marginally lower than the full-processor two-qubit
gate average of $1.85$\%. Readout errors for the selected qubits
($2.69$\% and $1.23$\%) also remained below the system-wide average
of $3.31$\%. The \textgreek{π}/2 rotation pulses (X/2 gates) were
implemented using optimized waveforms with system-normalized amplitudes
of 0.255 (Qubit 21) and 0.4451 (Qubit 26), corresponding to 40-ns
pulse durations. All parameters mentioned above are publicly accessible
through the Tianyan Quantum Cloud platform (Hardware parameters ,
e.g. $T_{1}$ , $T_{2}$, gate fidelities, are dynamically updated
on the publicly accessible calibration dashboard).

\begin{figure}
\includegraphics[scale=0.45]{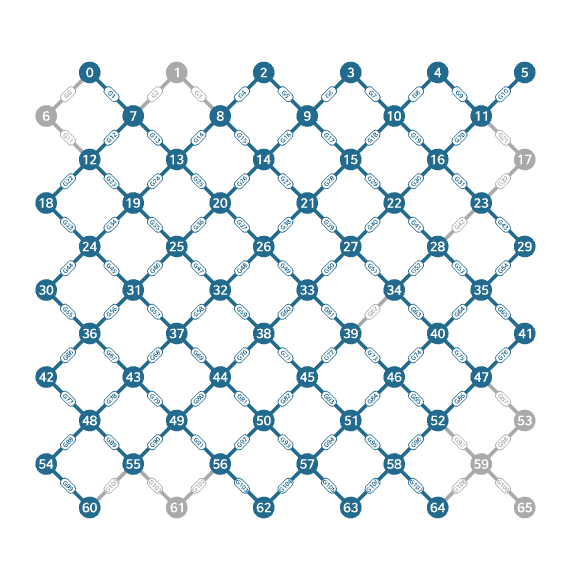}\caption{ Schematic topology of \textit{Zuchongzhi-2} superconducting quantum
processor. Gray-shaded qubits and couplers indicate disabled components
maybe due to calibration failures or coherence limitations. }\label{qubits}

\end{figure}

\section{ Circuit Transpilation Example}\label{Transpile}

\begin{figure*}
\begin{quantikz}
\lstick{$q_1$}&\gate{R_x(0.19743)}& &\targ{}&\gate{R_z(-0.31277)}&\gate{H}&\ctrl{1}&\gate{H}&\gate{R_x(\frac{\pi}{2})}&\\
\lstick{$q_0$}&\gate{R_x(0.98085)}&\gate{H}&\ctrl{-1}&\gate{H}&\gate{R_z(-0.2773)}&\targ{}&\gate{H}& &\\
\end{quantikz}
\caption{ The original circuit designed to compute the trace $\protect\tr{\left(Y_{1}\otimes X_{0}\right)}$.
The circuit includes the ansatz $\Gamma(\boldsymbol{\theta})$ with
the specified parameters for $t=0.5.$}\label{figtrans1}
\end{figure*}
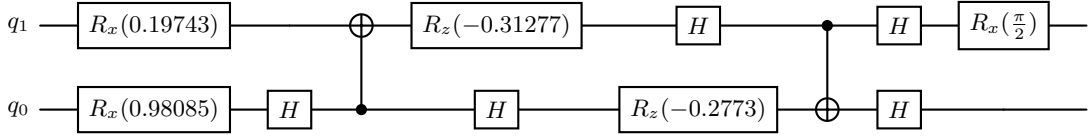

To illustrate the transpilation process, we provide an explicit example
for the circuit calculating $\tr{\left(Y_{1}\otimes X_{0}\right)}$.
The original circuit (Fig. \ref{figtrans1}) includes a parameterized
ansatz $\Gamma(\boldsymbol{\theta})=\Gamma(\theta_{0},\theta_{1},\theta_{2},\theta_{3})$
followed by gates for Pauli measurement. The corresponding parameters
shown in Fig. \ref{figtrans1} for $t=0.5$. While functionally correct,
this circuit uses generic gates unsuited for direct execution on the
Zuchongzhi-2 platform.

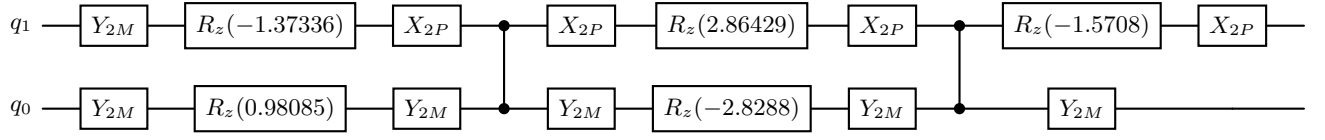
\begin{figure*}

\begin{quantikz}
\lstick{$q_1$}&\gate{Y_{2M}}&\gate{R_z(-1.37336)}&\gate{X_{2P}}&\control{}&\gate{X_{2P}}&\gate{R_z(2.86429)}&\gate{X_{2P}}&\control{}&\gate{R_z(-1.5708)}&\gate{X_{2P}}&\\
\lstick{$q_0$}&\gate{Y_{2M}}&\gate{R_z(0.98085)}&\gate{Y_{2M}}&\ctrl{-1}&\gate{Y_{2M}}&\gate{R_z(-2.8288)}&\gate{Y_{2M}}&\ctrl{-1}&\gate{Y_{2M}}& &\\
\end{quantikz}
\caption{ The transpiled version of the circuit for the quantum processor
hardware. The circuit is now in a format compatible with the basic
gate set, ensuring that no further decomposition is required on the
cloud server.}\label{figtrans2}
\end{figure*}

Using the \textit{bqskit} toolkit, we transpile the circuit into a
hardware-optimized form (Fig. \ref{figtrans2}). This process maps
all operations to the device’s native gate set $\{X2P,X2M,Y2P,Y2M,R_{z},CZ\}$
while preserving the logical functionality. Though the transpiled
circuit appears longer, it avoids further decomposition on the cloud
server of the quantum hardware, reducing error accumulation. The toolkit
occasionally generates different transpiled circuits, depending on
the optimization process. Researchers may further optimize gate sequences
using toolkit’s advanced compilation strategies, though the provided
version already ensures reliable execution.

\selectlanguage{english}%
\bibliographystyle{apsrev4-2}
\bibliography{reference}

\begin{thebibliography}{43}%
\makeatletter
\providecommand \@ifxundefined [1]{%
 \@ifx{#1\undefined}
}%
\providecommand \@ifnum [1]{%
 \ifnum #1\expandafter \@firstoftwo
 \else \expandafter \@secondoftwo
 \fi
}%
\providecommand \@ifx [1]{%
 \ifx #1\expandafter \@firstoftwo
 \else \expandafter \@secondoftwo
 \fi
}%
\providecommand \natexlab [1]{#1}%
\providecommand \enquote  [1]{``#1''}%
\providecommand \bibnamefont  [1]{#1}%
\providecommand \bibfnamefont [1]{#1}%
\providecommand \citenamefont [1]{#1}%
\providecommand \href@noop [0]{\@secondoftwo}%
\providecommand \href [0]{\begingroup \@sanitize@url \@href}%
\providecommand \@href[1]{\@@startlink{#1}\@@href}%
\providecommand \@@href[1]{\endgroup#1\@@endlink}%
\providecommand \@sanitize@url [0]{\catcode `\\12\catcode `\$12\catcode
  `\&12\catcode `\#12\catcode `\^12\catcode `\_12\catcode `\%12\relax}%
\providecommand \@@startlink[1]{}%
\providecommand \@@endlink[0]{}%
\providecommand \url  [0]{\begingroup\@sanitize@url \@url }%
\providecommand \@url [1]{\endgroup\@href {#1}{\urlprefix }}%
\providecommand \urlprefix  [0]{URL }%
\providecommand \Eprint [0]{\href }%
\providecommand \doibase [0]{https://doi.org/}%
\providecommand \selectlanguage [0]{\@gobble}%
\providecommand \bibinfo  [0]{\@secondoftwo}%
\providecommand \bibfield  [0]{\@secondoftwo}%
\providecommand \translation [1]{[#1]}%
\providecommand \BibitemOpen [0]{}%
\providecommand \bibitemStop [0]{}%
\providecommand \bibitemNoStop [0]{.\EOS\space}%
\providecommand \EOS [0]{\spacefactor3000\relax}%
\providecommand \BibitemShut  [1]{\csname bibitem#1\endcsname}%
\let\auto@bib@innerbib\@empty
%</preamble>
\bibitem [{\citenamefont {Nielsen}\ and\ \citenamefont
  {Chuang}(2010)}]{Nielsen2010}%
  \BibitemOpen
  \bibfield  {author} {\bibinfo {author} {\bibfnamefont {M.~A.}\ \bibnamefont
  {Nielsen}}\ and\ \bibinfo {author} {\bibfnamefont {I.~L.}\ \bibnamefont
  {Chuang}},\ }\href@noop {} {\emph {\bibinfo {title} {Quantum Computation and
  Quantum Information}}}\ (\bibinfo  {publisher} {Cambridge University Press},\
  \bibinfo {address} {Cambridge, England},\ \bibinfo {year} {2010})\BibitemShut
  {NoStop}%
\bibitem [{\citenamefont {Stanisic}\ \emph {et~al.}(2022)\citenamefont
  {Stanisic}, \citenamefont {Bosse}, \citenamefont {Gambetta}, \citenamefont
  {Santos}, \citenamefont {Mruczkiewicz}, \citenamefont {O’Brien},
  \citenamefont {Ostby},\ and\ \citenamefont {Montanaro}}]{Stanisic2022}%
  \BibitemOpen
  \bibfield  {author} {\bibinfo {author} {\bibfnamefont {S.}~\bibnamefont
  {Stanisic}}, \bibinfo {author} {\bibfnamefont {J.~L.}\ \bibnamefont {Bosse}},
  \bibinfo {author} {\bibfnamefont {F.~M.}\ \bibnamefont {Gambetta}}, \bibinfo
  {author} {\bibfnamefont {R.~A.}\ \bibnamefont {Santos}}, \bibinfo {author}
  {\bibfnamefont {W.}~\bibnamefont {Mruczkiewicz}}, \bibinfo {author}
  {\bibfnamefont {T.~E.}\ \bibnamefont {O’Brien}}, \bibinfo {author}
  {\bibfnamefont {E.}~\bibnamefont {Ostby}},\ and\ \bibinfo {author}
  {\bibfnamefont {A.}~\bibnamefont {Montanaro}},\ }\bibfield  {journal}
  {\bibinfo  {journal} {Nature Communications}\ }\textbf {\bibinfo {volume}
  {13}},\ \href {https://doi.org/10.1038/s41467-022-33335-4}
  {10.1038/s41467-022-33335-4} (\bibinfo {year} {2022})\BibitemShut {NoStop}%
\bibitem [{\citenamefont {Li}\ \emph {et~al.}(2024)\citenamefont {Li},
  \citenamefont {Yang}, \citenamefont {Lv}, \citenamefont {Qu}, \citenamefont
  {Wang}, \citenamefont {Sun},\ and\ \citenamefont {Ying}}]{Li_2024}%
  \BibitemOpen
  \bibfield  {author} {\bibinfo {author} {\bibfnamefont {H.}~\bibnamefont
  {Li}}, \bibinfo {author} {\bibfnamefont {Y.}~\bibnamefont {Yang}}, \bibinfo
  {author} {\bibfnamefont {P.}~\bibnamefont {Lv}}, \bibinfo {author}
  {\bibfnamefont {J.}~\bibnamefont {Qu}}, \bibinfo {author} {\bibfnamefont
  {Z.-H.}\ \bibnamefont {Wang}}, \bibinfo {author} {\bibfnamefont
  {J.}~\bibnamefont {Sun}},\ and\ \bibinfo {author} {\bibfnamefont
  {S.}~\bibnamefont {Ying}},\ }\href {https://doi.org/10.1088/1402-4896/ad770b}
  {\bibfield  {journal} {\bibinfo  {journal} {Physica Scripta}\ }\textbf
  {\bibinfo {volume} {99}},\ \bibinfo {pages} {105117} (\bibinfo {year}
  {2024})}\BibitemShut {NoStop}%
\bibitem [{\citenamefont {Santos}(2025)}]{PhysRevA.111.022618}%
  \BibitemOpen
  \bibfield  {author} {\bibinfo {author} {\bibfnamefont {A.~C.}\ \bibnamefont
  {Santos}},\ }\href {https://doi.org/10.1103/PhysRevA.111.022618} {\bibfield
  {journal} {\bibinfo  {journal} {Phys. Rev. A}\ }\textbf {\bibinfo {volume}
  {111}},\ \bibinfo {pages} {022618} (\bibinfo {year} {2025})}\BibitemShut
  {NoStop}%
\bibitem [{\citenamefont {Hu}\ \emph {et~al.}(2025)\citenamefont {Hu},
  \citenamefont {Xie}, \citenamefont {Poulsen}, \citenamefont {Zhou},
  \citenamefont {Chu}, \citenamefont {Liu}, \citenamefont {Zhou}, \citenamefont
  {Yuan}, \citenamefont {Shen}, \citenamefont {Liu}, \citenamefont {Zinner},
  \citenamefont {Tan}, \citenamefont {Santos},\ and\ \citenamefont
  {Yu}}]{Hu2025}%
  \BibitemOpen
  \bibfield  {author} {\bibinfo {author} {\bibfnamefont {C.-K.}\ \bibnamefont
  {Hu}}, \bibinfo {author} {\bibfnamefont {G.}~\bibnamefont {Xie}}, \bibinfo
  {author} {\bibfnamefont {K.}~\bibnamefont {Poulsen}}, \bibinfo {author}
  {\bibfnamefont {Y.}~\bibnamefont {Zhou}}, \bibinfo {author} {\bibfnamefont
  {J.}~\bibnamefont {Chu}}, \bibinfo {author} {\bibfnamefont {C.}~\bibnamefont
  {Liu}}, \bibinfo {author} {\bibfnamefont {R.}~\bibnamefont {Zhou}}, \bibinfo
  {author} {\bibfnamefont {H.}~\bibnamefont {Yuan}}, \bibinfo {author}
  {\bibfnamefont {Y.}~\bibnamefont {Shen}}, \bibinfo {author} {\bibfnamefont
  {S.}~\bibnamefont {Liu}}, \bibinfo {author} {\bibfnamefont {N.~T.}\
  \bibnamefont {Zinner}}, \bibinfo {author} {\bibfnamefont {D.}~\bibnamefont
  {Tan}}, \bibinfo {author} {\bibfnamefont {A.~C.}\ \bibnamefont {Santos}},\
  and\ \bibinfo {author} {\bibfnamefont {D.}~\bibnamefont {Yu}},\ }\bibfield
  {journal} {\bibinfo  {journal} {Nature Communications}\ }\textbf {\bibinfo
  {volume} {16}},\ \href {https://doi.org/10.1038/s41467-025-57812-8}
  {10.1038/s41467-025-57812-8} (\bibinfo {year} {2025})\BibitemShut {NoStop}%
\bibitem [{\citenamefont {Shor}(1994)}]{365700}%
  \BibitemOpen
  \bibfield  {author} {\bibinfo {author} {\bibfnamefont {P.}~\bibnamefont
  {Shor}},\ }in\ \href {https://doi.org/10.1109/SFCS.1994.365700} {\emph
  {\bibinfo {booktitle} {Proceedings 35th Annual Symposium on Foundations of
  Computer Science}}}\ (\bibinfo {year} {1994})\ pp.\ \bibinfo {pages}
  {124--134}\BibitemShut {NoStop}%
\bibitem [{\citenamefont {Shor}(1997)}]{Shor2}%
  \BibitemOpen
  \bibfield  {author} {\bibinfo {author} {\bibfnamefont {P.~W.}\ \bibnamefont
  {Shor}},\ }\href {https://doi.org/10.1137/S0097539795293172} {\bibfield
  {journal} {\bibinfo  {journal} {SIAM Journal on Computing}\ }\textbf
  {\bibinfo {volume} {26}},\ \bibinfo {pages} {1484} (\bibinfo {year}
  {1997})},\ \Eprint
  {https://arxiv.org/abs/https://doi.org/10.1137/S0097539795293172}
  {https://doi.org/10.1137/S0097539795293172} \BibitemShut {NoStop}%
\bibitem [{\citenamefont {Cao}\ \emph {et~al.}(2018)\citenamefont {Cao},
  \citenamefont {Romero},\ and\ \citenamefont {Aspuru-Guzik}}]{8585034}%
  \BibitemOpen
  \bibfield  {author} {\bibinfo {author} {\bibfnamefont {Y.}~\bibnamefont
  {Cao}}, \bibinfo {author} {\bibfnamefont {J.}~\bibnamefont {Romero}},\ and\
  \bibinfo {author} {\bibfnamefont {A.}~\bibnamefont {Aspuru-Guzik}},\ }\href
  {https://doi.org/10.1147/JRD.2018.2888987} {\bibfield  {journal} {\bibinfo
  {journal} {IBM Journal of Research and Development}\ }\textbf {\bibinfo
  {volume} {62}},\ \bibinfo {pages} {6:1} (\bibinfo {year} {2018})}\BibitemShut
  {NoStop}%
\bibitem [{\citenamefont {Steffen}\ \emph {et~al.}(2011)\citenamefont
  {Steffen}, \citenamefont {DiVincenzo}, \citenamefont {Chow}, \citenamefont
  {Theis},\ and\ \citenamefont {Ketchen}}]{SC1}%
  \BibitemOpen
  \bibfield  {author} {\bibinfo {author} {\bibfnamefont {M.}~\bibnamefont
  {Steffen}}, \bibinfo {author} {\bibfnamefont {D.~P.}\ \bibnamefont
  {DiVincenzo}}, \bibinfo {author} {\bibfnamefont {J.~M.}\ \bibnamefont
  {Chow}}, \bibinfo {author} {\bibfnamefont {T.~N.}\ \bibnamefont {Theis}},\
  and\ \bibinfo {author} {\bibfnamefont {M.~B.}\ \bibnamefont {Ketchen}},\
  }\href {https://doi.org/10.1147/JRD.2011.2165678} {\bibfield  {journal}
  {\bibinfo  {journal} {IBM Journal of Research and Development}\ }\textbf
  {\bibinfo {volume} {55}},\ \bibinfo {pages} {13:1} (\bibinfo {year}
  {2011})}\BibitemShut {NoStop}%
\bibitem [{\citenamefont {Kjaergaard}\ \emph {et~al.}(2020)\citenamefont
  {Kjaergaard}, \citenamefont {Schwartz}, \citenamefont {Braum\"{u}ller},
  \citenamefont {Krantz}, \citenamefont {Wang}, \citenamefont {Gustavsson},\
  and\ \citenamefont {Oliver}}]{SC_review}%
  \BibitemOpen
  \bibfield  {author} {\bibinfo {author} {\bibfnamefont {M.}~\bibnamefont
  {Kjaergaard}}, \bibinfo {author} {\bibfnamefont {M.~E.}\ \bibnamefont
  {Schwartz}}, \bibinfo {author} {\bibfnamefont {J.}~\bibnamefont
  {Braum\"{u}ller}}, \bibinfo {author} {\bibfnamefont {P.}~\bibnamefont
  {Krantz}}, \bibinfo {author} {\bibfnamefont {J.~I.-J.}\ \bibnamefont {Wang}},
  \bibinfo {author} {\bibfnamefont {S.}~\bibnamefont {Gustavsson}},\ and\
  \bibinfo {author} {\bibfnamefont {W.~D.}\ \bibnamefont {Oliver}},\ }\href
  {https://doi.org/10.1146/annurev-conmatphys-031119-050605} {\bibfield
  {journal} {\bibinfo  {journal} {Annual Review of Condensed Matter Physics}\
  }\textbf {\bibinfo {volume} {11}},\ \bibinfo {pages} {369} (\bibinfo {year}
  {2020})}\BibitemShut {NoStop}%
\bibitem [{\citenamefont {HAFFNER}\ \emph {et~al.}(2008)\citenamefont
  {HAFFNER}, \citenamefont {ROOS},\ and\ \citenamefont {BLATT}}]{ion1}%
  \BibitemOpen
  \bibfield  {author} {\bibinfo {author} {\bibfnamefont {H.}~\bibnamefont
  {HAFFNER}}, \bibinfo {author} {\bibfnamefont {C.}~\bibnamefont {ROOS}},\ and\
  \bibinfo {author} {\bibfnamefont {R.}~\bibnamefont {BLATT}},\ }\href
  {https://doi.org/10.1016/j.physrep.2008.09.003} {\bibfield  {journal}
  {\bibinfo  {journal} {Physics Reports}\ }\textbf {\bibinfo {volume} {469}},\
  \bibinfo {pages} {155} (\bibinfo {year} {2008})}\BibitemShut {NoStop}%
\bibitem [{\citenamefont {Bruzewicz}\ \emph {et~al.}(2019)\citenamefont
  {Bruzewicz}, \citenamefont {Chiaverini}, \citenamefont {McConnell},\ and\
  \citenamefont {Sage}}]{ion2}%
  \BibitemOpen
  \bibfield  {author} {\bibinfo {author} {\bibfnamefont {C.~D.}\ \bibnamefont
  {Bruzewicz}}, \bibinfo {author} {\bibfnamefont {J.}~\bibnamefont
  {Chiaverini}}, \bibinfo {author} {\bibfnamefont {R.}~\bibnamefont
  {McConnell}},\ and\ \bibinfo {author} {\bibfnamefont {J.~M.}\ \bibnamefont
  {Sage}},\ }\bibfield  {journal} {\bibinfo  {journal} {Applied Physics
  Reviews}\ }\textbf {\bibinfo {volume} {6}},\ \href
  {https://doi.org/10.1063/1.5088164} {10.1063/1.5088164} (\bibinfo {year}
  {2019})\BibitemShut {NoStop}%
\bibitem [{\citenamefont {Weiss}\ and\ \citenamefont
  {Saffman}(2017)}]{neutral1}%
  \BibitemOpen
  \bibfield  {author} {\bibinfo {author} {\bibfnamefont {D.~S.}\ \bibnamefont
  {Weiss}}\ and\ \bibinfo {author} {\bibfnamefont {M.}~\bibnamefont
  {Saffman}},\ }\href {https://doi.org/10.1063/pt.3.3626} {\bibfield  {journal}
  {\bibinfo  {journal} {Physics Today}\ }\textbf {\bibinfo {volume} {70}},\
  \bibinfo {pages} {44} (\bibinfo {year} {2017})}\BibitemShut {NoStop}%
\bibitem [{\citenamefont {Graham}\ \emph {et~al.}(2022)\citenamefont {Graham},
  \citenamefont {Song}, \citenamefont {Scott}, \citenamefont {Poole},
  \citenamefont {Phuttitarn}, \citenamefont {Jooya}, \citenamefont {Eichler},
  \citenamefont {Jiang}, \citenamefont {Marra}, \citenamefont {Grinkemeyer},
  \citenamefont {Kwon}, \citenamefont {Ebert}, \citenamefont {Cherek},
  \citenamefont {Lichtman}, \citenamefont {Gillette}, \citenamefont {Gilbert},
  \citenamefont {Bowman}, \citenamefont {Ballance}, \citenamefont {Campbell},
  \citenamefont {Dahl}, \citenamefont {Crawford}, \citenamefont {Blunt},
  \citenamefont {Rogers}, \citenamefont {Noel},\ and\ \citenamefont
  {Saffman}}]{neutral2}%
  \BibitemOpen
  \bibfield  {author} {\bibinfo {author} {\bibfnamefont {T.~M.}\ \bibnamefont
  {Graham}}, \bibinfo {author} {\bibfnamefont {Y.}~\bibnamefont {Song}},
  \bibinfo {author} {\bibfnamefont {J.}~\bibnamefont {Scott}}, \bibinfo
  {author} {\bibfnamefont {C.}~\bibnamefont {Poole}}, \bibinfo {author}
  {\bibfnamefont {L.}~\bibnamefont {Phuttitarn}}, \bibinfo {author}
  {\bibfnamefont {K.}~\bibnamefont {Jooya}}, \bibinfo {author} {\bibfnamefont
  {P.}~\bibnamefont {Eichler}}, \bibinfo {author} {\bibfnamefont
  {X.}~\bibnamefont {Jiang}}, \bibinfo {author} {\bibfnamefont
  {A.}~\bibnamefont {Marra}}, \bibinfo {author} {\bibfnamefont
  {B.}~\bibnamefont {Grinkemeyer}}, \bibinfo {author} {\bibfnamefont
  {M.}~\bibnamefont {Kwon}}, \bibinfo {author} {\bibfnamefont {M.}~\bibnamefont
  {Ebert}}, \bibinfo {author} {\bibfnamefont {J.}~\bibnamefont {Cherek}},
  \bibinfo {author} {\bibfnamefont {M.~T.}\ \bibnamefont {Lichtman}}, \bibinfo
  {author} {\bibfnamefont {M.}~\bibnamefont {Gillette}}, \bibinfo {author}
  {\bibfnamefont {J.}~\bibnamefont {Gilbert}}, \bibinfo {author} {\bibfnamefont
  {D.}~\bibnamefont {Bowman}}, \bibinfo {author} {\bibfnamefont
  {T.}~\bibnamefont {Ballance}}, \bibinfo {author} {\bibfnamefont
  {C.}~\bibnamefont {Campbell}}, \bibinfo {author} {\bibfnamefont {E.~D.}\
  \bibnamefont {Dahl}}, \bibinfo {author} {\bibfnamefont {O.}~\bibnamefont
  {Crawford}}, \bibinfo {author} {\bibfnamefont {N.~S.}\ \bibnamefont {Blunt}},
  \bibinfo {author} {\bibfnamefont {B.}~\bibnamefont {Rogers}}, \bibinfo
  {author} {\bibfnamefont {T.}~\bibnamefont {Noel}},\ and\ \bibinfo {author}
  {\bibfnamefont {M.}~\bibnamefont {Saffman}},\ }\href
  {https://doi.org/10.1038/s41586-022-04603-6} {\bibfield  {journal} {\bibinfo
  {journal} {Nature}\ }\textbf {\bibinfo {volume} {604}},\ \bibinfo {pages}
  {457} (\bibinfo {year} {2022})}\BibitemShut {NoStop}%
\bibitem [{\citenamefont {Knill}\ \emph {et~al.}(2001)\citenamefont {Knill},
  \citenamefont {Laflamme},\ and\ \citenamefont {Milburn}}]{KLM}%
  \BibitemOpen
  \bibfield  {author} {\bibinfo {author} {\bibfnamefont {E.}~\bibnamefont
  {Knill}}, \bibinfo {author} {\bibfnamefont {R.}~\bibnamefont {Laflamme}},\
  and\ \bibinfo {author} {\bibfnamefont {G.~J.}\ \bibnamefont {Milburn}},\
  }\href {https://doi.org/10.1038/35051009} {\bibfield  {journal} {\bibinfo
  {journal} {Nature}\ }\textbf {\bibinfo {volume} {409}},\ \bibinfo {pages}
  {46} (\bibinfo {year} {2001})}\BibitemShut {NoStop}%
\bibitem [{\citenamefont {Raussendorf}\ and\ \citenamefont
  {Briegel}(2001)}]{MBQC}%
  \BibitemOpen
  \bibfield  {author} {\bibinfo {author} {\bibfnamefont {R.}~\bibnamefont
  {Raussendorf}}\ and\ \bibinfo {author} {\bibfnamefont {H.~J.}\ \bibnamefont
  {Briegel}},\ }\href {https://doi.org/10.1103/physrevlett.86.5188} {\bibfield
  {journal} {\bibinfo  {journal} {Physical Review Letters}\ }\textbf {\bibinfo
  {volume} {86}},\ \bibinfo {pages} {5188} (\bibinfo {year}
  {2001})}\BibitemShut {NoStop}%
\bibitem [{\citenamefont {Bartolucci}\ \emph {et~al.}(2023)\citenamefont
  {Bartolucci}, \citenamefont {Birchall}, \citenamefont {Bomb{\'\i}n},
  \citenamefont {Cable}, \citenamefont {Dawson}, \citenamefont
  {Gimeno-Segovia}, \citenamefont {Johnston}, \citenamefont {Kieling},
  \citenamefont {Nickerson}, \citenamefont {Pant}, \citenamefont {Pastawski},
  \citenamefont {Rudolph},\ and\ \citenamefont {Sparrow}}]{FBQC1}%
  \BibitemOpen
  \bibfield  {author} {\bibinfo {author} {\bibfnamefont {S.}~\bibnamefont
  {Bartolucci}}, \bibinfo {author} {\bibfnamefont {P.}~\bibnamefont
  {Birchall}}, \bibinfo {author} {\bibfnamefont {H.}~\bibnamefont
  {Bomb{\'\i}n}}, \bibinfo {author} {\bibfnamefont {H.}~\bibnamefont {Cable}},
  \bibinfo {author} {\bibfnamefont {C.}~\bibnamefont {Dawson}}, \bibinfo
  {author} {\bibfnamefont {M.}~\bibnamefont {Gimeno-Segovia}}, \bibinfo
  {author} {\bibfnamefont {E.}~\bibnamefont {Johnston}}, \bibinfo {author}
  {\bibfnamefont {K.}~\bibnamefont {Kieling}}, \bibinfo {author} {\bibfnamefont
  {N.}~\bibnamefont {Nickerson}}, \bibinfo {author} {\bibfnamefont
  {M.}~\bibnamefont {Pant}}, \bibinfo {author} {\bibfnamefont {F.}~\bibnamefont
  {Pastawski}}, \bibinfo {author} {\bibfnamefont {T.}~\bibnamefont {Rudolph}},\
  and\ \bibinfo {author} {\bibfnamefont {C.}~\bibnamefont {Sparrow}},\
  }\href@noop {} {\bibfield  {journal} {\bibinfo  {journal} {Nat. Commun.}\
  }\textbf {\bibinfo {volume} {14}},\ \bibinfo {pages} {912} (\bibinfo {year}
  {2023})}\BibitemShut {NoStop}%
\bibitem [{\citenamefont {Bourassa}\ \emph {et~al.}(2021)\citenamefont
  {Bourassa}, \citenamefont {Alexander}, \citenamefont {Vasmer}, \citenamefont
  {Patil}, \citenamefont {Tzitrin}, \citenamefont {Matsuura}, \citenamefont
  {Su}, \citenamefont {Baragiola}, \citenamefont {Guha}, \citenamefont
  {Dauphinais}, \citenamefont {Sabapathy}, \citenamefont {Menicucci},\ and\
  \citenamefont {Dhand}}]{Xanadu1}%
  \BibitemOpen
  \bibfield  {author} {\bibinfo {author} {\bibfnamefont {J.~E.}\ \bibnamefont
  {Bourassa}}, \bibinfo {author} {\bibfnamefont {R.~N.}\ \bibnamefont
  {Alexander}}, \bibinfo {author} {\bibfnamefont {M.}~\bibnamefont {Vasmer}},
  \bibinfo {author} {\bibfnamefont {A.}~\bibnamefont {Patil}}, \bibinfo
  {author} {\bibfnamefont {I.}~\bibnamefont {Tzitrin}}, \bibinfo {author}
  {\bibfnamefont {T.}~\bibnamefont {Matsuura}}, \bibinfo {author}
  {\bibfnamefont {D.}~\bibnamefont {Su}}, \bibinfo {author} {\bibfnamefont
  {B.~Q.}\ \bibnamefont {Baragiola}}, \bibinfo {author} {\bibfnamefont
  {S.}~\bibnamefont {Guha}}, \bibinfo {author} {\bibfnamefont {G.}~\bibnamefont
  {Dauphinais}}, \bibinfo {author} {\bibfnamefont {K.~K.}\ \bibnamefont
  {Sabapathy}}, \bibinfo {author} {\bibfnamefont {N.~C.}\ \bibnamefont
  {Menicucci}},\ and\ \bibinfo {author} {\bibfnamefont {I.}~\bibnamefont
  {Dhand}},\ }\href {https://doi.org/10.22331/q-2021-02-04-392} {\bibfield
  {journal} {\bibinfo  {journal} {Quantum}\ }\textbf {\bibinfo {volume} {5}},\
  \bibinfo {pages} {392} (\bibinfo {year} {2021})}\BibitemShut {NoStop}%
\bibitem [{\citenamefont {Pant}\ \emph {et~al.}(2019)\citenamefont {Pant},
  \citenamefont {Towsley}, \citenamefont {Englund},\ and\ \citenamefont
  {Guha}}]{percolation}%
  \BibitemOpen
  \bibfield  {author} {\bibinfo {author} {\bibfnamefont {M.}~\bibnamefont
  {Pant}}, \bibinfo {author} {\bibfnamefont {D.}~\bibnamefont {Towsley}},
  \bibinfo {author} {\bibfnamefont {D.}~\bibnamefont {Englund}},\ and\ \bibinfo
  {author} {\bibfnamefont {S.}~\bibnamefont {Guha}},\ }\bibfield  {journal}
  {\bibinfo  {journal} {Nature Communications}\ }\textbf {\bibinfo {volume}
  {10}},\ \href {https://doi.org/10.1038/s41467-019-08948-x}
  {10.1038/s41467-019-08948-x} (\bibinfo {year} {2019})\BibitemShut {NoStop}%
\bibitem [{\citenamefont {Gottesman}\ \emph {et~al.}(2001)\citenamefont
  {Gottesman}, \citenamefont {Kitaev},\ and\ \citenamefont {Preskill}}]{GKP}%
  \BibitemOpen
  \bibfield  {author} {\bibinfo {author} {\bibfnamefont {D.}~\bibnamefont
  {Gottesman}}, \bibinfo {author} {\bibfnamefont {A.}~\bibnamefont {Kitaev}},\
  and\ \bibinfo {author} {\bibfnamefont {J.}~\bibnamefont {Preskill}},\ }\href
  {https://doi.org/10.1103/PhysRevA.64.012310} {\bibfield  {journal} {\bibinfo
  {journal} {Phys. Rev. A}\ }\textbf {\bibinfo {volume} {64}},\ \bibinfo
  {pages} {012310} (\bibinfo {year} {2001})}\BibitemShut {NoStop}%
\bibitem [{\citenamefont {Somma}\ \emph {et~al.}(2003)\citenamefont {Somma},
  \citenamefont {Ortiz}, \citenamefont {Knill},\ and\ \citenamefont
  {Gubernatis}}]{onehot1}%
  \BibitemOpen
  \bibfield  {author} {\bibinfo {author} {\bibfnamefont {R.~D.}\ \bibnamefont
  {Somma}}, \bibinfo {author} {\bibfnamefont {G.}~\bibnamefont {Ortiz}},
  \bibinfo {author} {\bibfnamefont {E.~H.}\ \bibnamefont {Knill}},\ and\
  \bibinfo {author} {\bibfnamefont {J.}~\bibnamefont {Gubernatis}},\ }in\ \href
  {https://doi.org/10.1117/12.487249} {\emph {\bibinfo {booktitle} {Quantum
  Information and Computation}}},\ \bibinfo {editor} {edited by\ \bibinfo
  {editor} {\bibfnamefont {E.}~\bibnamefont {Donkor}}, \bibinfo {editor}
  {\bibfnamefont {A.~R.}\ \bibnamefont {Pirich}},\ and\ \bibinfo {editor}
  {\bibfnamefont {H.~E.}\ \bibnamefont {Brandt}}}\ (\bibinfo  {publisher}
  {SPIE},\ \bibinfo {year} {2003})\BibitemShut {NoStop}%
\bibitem [{\citenamefont {Di~Matteo}\ \emph {et~al.}(2021)\citenamefont
  {Di~Matteo}, \citenamefont {McCoy}, \citenamefont {Gysbers}, \citenamefont
  {Miyagi}, \citenamefont {Woloshyn},\ and\ \citenamefont
  {Navr\'atil}}]{graycode1}%
  \BibitemOpen
  \bibfield  {author} {\bibinfo {author} {\bibfnamefont {O.}~\bibnamefont
  {Di~Matteo}}, \bibinfo {author} {\bibfnamefont {A.}~\bibnamefont {McCoy}},
  \bibinfo {author} {\bibfnamefont {P.}~\bibnamefont {Gysbers}}, \bibinfo
  {author} {\bibfnamefont {T.}~\bibnamefont {Miyagi}}, \bibinfo {author}
  {\bibfnamefont {R.~M.}\ \bibnamefont {Woloshyn}},\ and\ \bibinfo {author}
  {\bibfnamefont {P.}~\bibnamefont {Navr\'atil}},\ }\href
  {https://doi.org/10.1103/PhysRevA.103.042405} {\bibfield  {journal} {\bibinfo
   {journal} {Phys. Rev. A}\ }\textbf {\bibinfo {volume} {103}},\ \bibinfo
  {pages} {042405} (\bibinfo {year} {2021})}\BibitemShut {NoStop}%
\bibitem [{\citenamefont {Sawaya}\ \emph {et~al.}(2020)\citenamefont {Sawaya},
  \citenamefont {Menke}, \citenamefont {Kyaw}, \citenamefont {Johri},
  \citenamefont {Aspuru-Guzik},\ and\ \citenamefont {Guerreschi}}]{Sawaya2020}%
  \BibitemOpen
  \bibfield  {author} {\bibinfo {author} {\bibfnamefont {N.~P.~D.}\
  \bibnamefont {Sawaya}}, \bibinfo {author} {\bibfnamefont {T.}~\bibnamefont
  {Menke}}, \bibinfo {author} {\bibfnamefont {T.~H.}\ \bibnamefont {Kyaw}},
  \bibinfo {author} {\bibfnamefont {S.}~\bibnamefont {Johri}}, \bibinfo
  {author} {\bibfnamefont {A.}~\bibnamefont {Aspuru-Guzik}},\ and\ \bibinfo
  {author} {\bibfnamefont {G.~G.}\ \bibnamefont {Guerreschi}},\ }\bibfield
  {journal} {\bibinfo  {journal} {npj Quantum Information}\ }\textbf {\bibinfo
  {volume} {6}},\ \href {https://doi.org/10.1038/s41534-020-0278-0}
  {10.1038/s41534-020-0278-0} (\bibinfo {year} {2020})\BibitemShut {NoStop}%
\bibitem [{\citenamefont {Kottmann}\ \emph {et~al.}(2021)\citenamefont
  {Kottmann}, \citenamefont {Krenn}, \citenamefont {Kyaw}, \citenamefont
  {Alperin-Lea},\ and\ \citenamefont {Aspuru-Guzik}}]{Kottmann2021}%
  \BibitemOpen
  \bibfield  {author} {\bibinfo {author} {\bibfnamefont {J.~S.}\ \bibnamefont
  {Kottmann}}, \bibinfo {author} {\bibfnamefont {M.}~\bibnamefont {Krenn}},
  \bibinfo {author} {\bibfnamefont {T.~H.}\ \bibnamefont {Kyaw}}, \bibinfo
  {author} {\bibfnamefont {S.}~\bibnamefont {Alperin-Lea}},\ and\ \bibinfo
  {author} {\bibfnamefont {A.}~\bibnamefont {Aspuru-Guzik}},\ }\href
  {https://doi.org/10.1088/2058-9565/abfc94} {\bibfield  {journal} {\bibinfo
  {journal} {Quantum Science and Technology}\ }\textbf {\bibinfo {volume}
  {6}},\ \bibinfo {pages} {035010} (\bibinfo {year} {2021})}\BibitemShut
  {NoStop}%
\bibitem [{\citenamefont {Chin}\ \emph {et~al.}(2024)\citenamefont {Chin},
  \citenamefont {Kim},\ and\ \citenamefont {Huh}}]{Chin2024}%
  \BibitemOpen
  \bibfield  {author} {\bibinfo {author} {\bibfnamefont {S.}~\bibnamefont
  {Chin}}, \bibinfo {author} {\bibfnamefont {J.}~\bibnamefont {Kim}},\ and\
  \bibinfo {author} {\bibfnamefont {J.}~\bibnamefont {Huh}},\ }\bibfield
  {journal} {\bibinfo  {journal} {SciPost Physics Core}\ }\textbf {\bibinfo
  {volume} {7}},\ \href {https://doi.org/10.21468/scipostphyscore.7.3.042}
  {10.21468/scipostphyscore.7.3.042} (\bibinfo {year} {2024})\BibitemShut
  {NoStop}%
\bibitem [{\citenamefont {Encinar}\ \emph {et~al.}(2021)\citenamefont
  {Encinar}, \citenamefont {Agust\'{\i}},\ and\ \citenamefont
  {Sab\'{\i}n}}]{squeeze1}%
  \BibitemOpen
  \bibfield  {author} {\bibinfo {author} {\bibfnamefont {P.~C.}\ \bibnamefont
  {Encinar}}, \bibinfo {author} {\bibfnamefont {A.}~\bibnamefont
  {Agust\'{\i}}},\ and\ \bibinfo {author} {\bibfnamefont {C.}~\bibnamefont
  {Sab\'{\i}n}},\ }\href {https://doi.org/10.1103/PhysRevA.104.052609}
  {\bibfield  {journal} {\bibinfo  {journal} {Phys. Rev. A}\ }\textbf {\bibinfo
  {volume} {104}},\ \bibinfo {pages} {052609} (\bibinfo {year}
  {2021})}\BibitemShut {NoStop}%
\bibitem [{\citenamefont {Li}\ and\ \citenamefont {Benjamin}(2017)}]{VQS1}%
  \BibitemOpen
  \bibfield  {author} {\bibinfo {author} {\bibfnamefont {Y.}~\bibnamefont
  {Li}}\ and\ \bibinfo {author} {\bibfnamefont {S.~C.}\ \bibnamefont
  {Benjamin}},\ }\href {https://doi.org/10.1103/PhysRevX.7.021050} {\bibfield
  {journal} {\bibinfo  {journal} {Phys. Rev. X}\ }\textbf {\bibinfo {volume}
  {7}},\ \bibinfo {pages} {021050} (\bibinfo {year} {2017})}\BibitemShut
  {NoStop}%
\bibitem [{\citenamefont {Endo}\ \emph {et~al.}(2020)\citenamefont {Endo},
  \citenamefont {Kurata},\ and\ \citenamefont {Nakagawa}}]{VQS2}%
  \BibitemOpen
  \bibfield  {author} {\bibinfo {author} {\bibfnamefont {S.}~\bibnamefont
  {Endo}}, \bibinfo {author} {\bibfnamefont {I.}~\bibnamefont {Kurata}},\ and\
  \bibinfo {author} {\bibfnamefont {Y.~O.}\ \bibnamefont {Nakagawa}},\ }\href
  {https://doi.org/10.1103/PhysRevResearch.2.033281} {\bibfield  {journal}
  {\bibinfo  {journal} {Phys. Rev. Res.}\ }\textbf {\bibinfo {volume} {2}},\
  \bibinfo {pages} {033281} (\bibinfo {year} {2020})}\BibitemShut {NoStop}%
\bibitem [{\citenamefont {Libbi}\ \emph {et~al.}(2022)\citenamefont {Libbi},
  \citenamefont {Rizzo}, \citenamefont {Tacchino}, \citenamefont {Marzari},\
  and\ \citenamefont {Tavernelli}}]{VQS3}%
  \BibitemOpen
  \bibfield  {author} {\bibinfo {author} {\bibfnamefont {F.}~\bibnamefont
  {Libbi}}, \bibinfo {author} {\bibfnamefont {J.}~\bibnamefont {Rizzo}},
  \bibinfo {author} {\bibfnamefont {F.}~\bibnamefont {Tacchino}}, \bibinfo
  {author} {\bibfnamefont {N.}~\bibnamefont {Marzari}},\ and\ \bibinfo {author}
  {\bibfnamefont {I.}~\bibnamefont {Tavernelli}},\ }\href
  {https://doi.org/10.1103/PhysRevResearch.4.043038} {\bibfield  {journal}
  {\bibinfo  {journal} {Phys. Rev. Res.}\ }\textbf {\bibinfo {volume} {4}},\
  \bibinfo {pages} {043038} (\bibinfo {year} {2022})}\BibitemShut {NoStop}%
\bibitem [{\citenamefont {Suzuki}(1976)}]{Suzuki}%
  \BibitemOpen
  \bibfield  {author} {\bibinfo {author} {\bibfnamefont {M.}~\bibnamefont
  {Suzuki}},\ }\href {https://doi.org/10.1007/bf01609348} {\bibfield  {journal}
  {\bibinfo  {journal} {Communications in Mathematical Physics}\ }\textbf
  {\bibinfo {volume} {51}},\ \bibinfo {pages} {183} (\bibinfo {year}
  {1976})}\BibitemShut {NoStop}%
\bibitem [{\citenamefont {Wu}\ \emph {et~al.}(2021)\citenamefont {Wu},
  \citenamefont {Bao}, \citenamefont {Cao}, \citenamefont {Chen}, \citenamefont
  {Chen}, \citenamefont {Chen}, \citenamefont {Chung}, \citenamefont {Deng},
  \citenamefont {Du}, \citenamefont {Fan}, \citenamefont {Gong}, \citenamefont
  {Guo}, \citenamefont {Guo}, \citenamefont {Guo}, \citenamefont {Han},
  \citenamefont {Hong}, \citenamefont {Huang}, \citenamefont {Huo},
  \citenamefont {Li}, \citenamefont {Li}, \citenamefont {Li}, \citenamefont
  {Li}, \citenamefont {Liang}, \citenamefont {Lin}, \citenamefont {Lin},
  \citenamefont {Qian}, \citenamefont {Qiao}, \citenamefont {Rong},
  \citenamefont {Su}, \citenamefont {Sun}, \citenamefont {Wang}, \citenamefont
  {Wang}, \citenamefont {Wu}, \citenamefont {Xu}, \citenamefont {Yan},
  \citenamefont {Yang}, \citenamefont {Yang}, \citenamefont {Ye}, \citenamefont
  {Yin}, \citenamefont {Ying}, \citenamefont {Yu}, \citenamefont {Zha},
  \citenamefont {Zhang}, \citenamefont {Zhang}, \citenamefont {Zhang},
  \citenamefont {Zhang}, \citenamefont {Zhao}, \citenamefont {Zhao},
  \citenamefont {Zhou}, \citenamefont {Zhu}, \citenamefont {Lu}, \citenamefont
  {Peng}, \citenamefont {Zhu},\ and\ \citenamefont {Pan}}]{guodun_yingjian}%
  \BibitemOpen
  \bibfield  {author} {\bibinfo {author} {\bibfnamefont {Y.}~\bibnamefont
  {Wu}}, \bibinfo {author} {\bibfnamefont {W.-S.}\ \bibnamefont {Bao}},
  \bibinfo {author} {\bibfnamefont {S.}~\bibnamefont {Cao}}, \bibinfo {author}
  {\bibfnamefont {F.}~\bibnamefont {Chen}}, \bibinfo {author} {\bibfnamefont
  {M.-C.}\ \bibnamefont {Chen}}, \bibinfo {author} {\bibfnamefont
  {X.}~\bibnamefont {Chen}}, \bibinfo {author} {\bibfnamefont {T.-H.}\
  \bibnamefont {Chung}}, \bibinfo {author} {\bibfnamefont {H.}~\bibnamefont
  {Deng}}, \bibinfo {author} {\bibfnamefont {Y.}~\bibnamefont {Du}}, \bibinfo
  {author} {\bibfnamefont {D.}~\bibnamefont {Fan}}, \bibinfo {author}
  {\bibfnamefont {M.}~\bibnamefont {Gong}}, \bibinfo {author} {\bibfnamefont
  {C.}~\bibnamefont {Guo}}, \bibinfo {author} {\bibfnamefont {C.}~\bibnamefont
  {Guo}}, \bibinfo {author} {\bibfnamefont {S.}~\bibnamefont {Guo}}, \bibinfo
  {author} {\bibfnamefont {L.}~\bibnamefont {Han}}, \bibinfo {author}
  {\bibfnamefont {L.}~\bibnamefont {Hong}}, \bibinfo {author} {\bibfnamefont
  {H.-L.}\ \bibnamefont {Huang}}, \bibinfo {author} {\bibfnamefont {Y.-H.}\
  \bibnamefont {Huo}}, \bibinfo {author} {\bibfnamefont {L.}~\bibnamefont
  {Li}}, \bibinfo {author} {\bibfnamefont {N.}~\bibnamefont {Li}}, \bibinfo
  {author} {\bibfnamefont {S.}~\bibnamefont {Li}}, \bibinfo {author}
  {\bibfnamefont {Y.}~\bibnamefont {Li}}, \bibinfo {author} {\bibfnamefont
  {F.}~\bibnamefont {Liang}}, \bibinfo {author} {\bibfnamefont
  {C.}~\bibnamefont {Lin}}, \bibinfo {author} {\bibfnamefont {J.}~\bibnamefont
  {Lin}}, \bibinfo {author} {\bibfnamefont {H.}~\bibnamefont {Qian}}, \bibinfo
  {author} {\bibfnamefont {D.}~\bibnamefont {Qiao}}, \bibinfo {author}
  {\bibfnamefont {H.}~\bibnamefont {Rong}}, \bibinfo {author} {\bibfnamefont
  {H.}~\bibnamefont {Su}}, \bibinfo {author} {\bibfnamefont {L.}~\bibnamefont
  {Sun}}, \bibinfo {author} {\bibfnamefont {L.}~\bibnamefont {Wang}}, \bibinfo
  {author} {\bibfnamefont {S.}~\bibnamefont {Wang}}, \bibinfo {author}
  {\bibfnamefont {D.}~\bibnamefont {Wu}}, \bibinfo {author} {\bibfnamefont
  {Y.}~\bibnamefont {Xu}}, \bibinfo {author} {\bibfnamefont {K.}~\bibnamefont
  {Yan}}, \bibinfo {author} {\bibfnamefont {W.}~\bibnamefont {Yang}}, \bibinfo
  {author} {\bibfnamefont {Y.}~\bibnamefont {Yang}}, \bibinfo {author}
  {\bibfnamefont {Y.}~\bibnamefont {Ye}}, \bibinfo {author} {\bibfnamefont
  {J.}~\bibnamefont {Yin}}, \bibinfo {author} {\bibfnamefont {C.}~\bibnamefont
  {Ying}}, \bibinfo {author} {\bibfnamefont {J.}~\bibnamefont {Yu}}, \bibinfo
  {author} {\bibfnamefont {C.}~\bibnamefont {Zha}}, \bibinfo {author}
  {\bibfnamefont {C.}~\bibnamefont {Zhang}}, \bibinfo {author} {\bibfnamefont
  {H.}~\bibnamefont {Zhang}}, \bibinfo {author} {\bibfnamefont
  {K.}~\bibnamefont {Zhang}}, \bibinfo {author} {\bibfnamefont
  {Y.}~\bibnamefont {Zhang}}, \bibinfo {author} {\bibfnamefont
  {H.}~\bibnamefont {Zhao}}, \bibinfo {author} {\bibfnamefont {Y.}~\bibnamefont
  {Zhao}}, \bibinfo {author} {\bibfnamefont {L.}~\bibnamefont {Zhou}}, \bibinfo
  {author} {\bibfnamefont {Q.}~\bibnamefont {Zhu}}, \bibinfo {author}
  {\bibfnamefont {C.-Y.}\ \bibnamefont {Lu}}, \bibinfo {author} {\bibfnamefont
  {C.-Z.}\ \bibnamefont {Peng}}, \bibinfo {author} {\bibfnamefont
  {X.}~\bibnamefont {Zhu}},\ and\ \bibinfo {author} {\bibfnamefont {J.-W.}\
  \bibnamefont {Pan}},\ }\href {https://doi.org/10.1103/PhysRevLett.127.180501}
  {\bibfield  {journal} {\bibinfo  {journal} {Phys. Rev. Lett.}\ }\textbf
  {\bibinfo {volume} {127}},\ \bibinfo {pages} {180501} (\bibinfo {year}
  {2021})}\BibitemShut {NoStop}%
\bibitem [{\citenamefont {Vogel}\ and\ \citenamefont {Welsch}(2006)}]{Vogel}%
  \BibitemOpen
  \bibfield  {author} {\bibinfo {author} {\bibfnamefont {W.}~\bibnamefont
  {Vogel}}\ and\ \bibinfo {author} {\bibfnamefont {D.-G.}\ \bibnamefont
  {Welsch}},\ }\href {https://doi.org/10.1002/3527608524} {\emph {\bibinfo
  {title} {Quantum Optics}}}\ (\bibinfo  {publisher} {Wiley},\ \bibinfo {year}
  {2006})\BibitemShut {NoStop}%
\bibitem [{\citenamefont {Peruzzo}\ \emph {et~al.}(2014)\citenamefont
  {Peruzzo}, \citenamefont {McClean}, \citenamefont {Shadbolt}, \citenamefont
  {Yung}, \citenamefont {Zhou}, \citenamefont {Love}, \citenamefont
  {Aspuru-Guzik},\ and\ \citenamefont {O'Brien}}]{VQE_first}%
  \BibitemOpen
  \bibfield  {author} {\bibinfo {author} {\bibfnamefont {A.}~\bibnamefont
  {Peruzzo}}, \bibinfo {author} {\bibfnamefont {J.}~\bibnamefont {McClean}},
  \bibinfo {author} {\bibfnamefont {P.}~\bibnamefont {Shadbolt}}, \bibinfo
  {author} {\bibfnamefont {M.-H.}\ \bibnamefont {Yung}}, \bibinfo {author}
  {\bibfnamefont {X.-Q.}\ \bibnamefont {Zhou}}, \bibinfo {author}
  {\bibfnamefont {P.~J.}\ \bibnamefont {Love}}, \bibinfo {author}
  {\bibfnamefont {A.}~\bibnamefont {Aspuru-Guzik}},\ and\ \bibinfo {author}
  {\bibfnamefont {J.~L.}\ \bibnamefont {O'Brien}},\ }\bibfield  {journal}
  {\bibinfo  {journal} {Nature Communications}\ }\textbf {\bibinfo {volume}
  {5}},\ \href {https://doi.org/10.1038/ncomms5213} {10.1038/ncomms5213}
  (\bibinfo {year} {2014})\BibitemShut {NoStop}%
\bibitem [{\citenamefont {Tilly}\ \emph {et~al.}(2022)\citenamefont {Tilly},
  \citenamefont {Chen}, \citenamefont {Cao}, \citenamefont {Picozzi},
  \citenamefont {Setia}, \citenamefont {Li}, \citenamefont {Grant},
  \citenamefont {Wossnig}, \citenamefont {Rungger}, \citenamefont {Booth},\
  and\ \citenamefont {Tennyson}}]{VQE_first_review}%
  \BibitemOpen
  \bibfield  {author} {\bibinfo {author} {\bibfnamefont {J.}~\bibnamefont
  {Tilly}}, \bibinfo {author} {\bibfnamefont {H.}~\bibnamefont {Chen}},
  \bibinfo {author} {\bibfnamefont {S.}~\bibnamefont {Cao}}, \bibinfo {author}
  {\bibfnamefont {D.}~\bibnamefont {Picozzi}}, \bibinfo {author} {\bibfnamefont
  {K.}~\bibnamefont {Setia}}, \bibinfo {author} {\bibfnamefont
  {Y.}~\bibnamefont {Li}}, \bibinfo {author} {\bibfnamefont {E.}~\bibnamefont
  {Grant}}, \bibinfo {author} {\bibfnamefont {L.}~\bibnamefont {Wossnig}},
  \bibinfo {author} {\bibfnamefont {I.}~\bibnamefont {Rungger}}, \bibinfo
  {author} {\bibfnamefont {G.~H.}\ \bibnamefont {Booth}},\ and\ \bibinfo
  {author} {\bibfnamefont {J.}~\bibnamefont {Tennyson}},\ }\href
  {https://doi.org/https://doi.org/10.1016/j.physrep.2022.08.003} {\bibfield
  {journal} {\bibinfo  {journal} {Physics Reports}\ }\textbf {\bibinfo {volume}
  {986}},\ \bibinfo {pages} {1} (\bibinfo {year} {2022})},\ \bibinfo {note}
  {the Variational Quantum Eigensolver: a review of methods and best
  practices}\BibitemShut {NoStop}%
\bibitem [{\citenamefont {Broeckhove}\ \emph {et~al.}(1988)\citenamefont
  {Broeckhove}, \citenamefont {Lathouwers}, \citenamefont {Kesteloot},\ and\
  \citenamefont {{Van Leuven}}}]{BROECKHOVE1988547}%
  \BibitemOpen
  \bibfield  {author} {\bibinfo {author} {\bibfnamefont {J.}~\bibnamefont
  {Broeckhove}}, \bibinfo {author} {\bibfnamefont {L.}~\bibnamefont
  {Lathouwers}}, \bibinfo {author} {\bibfnamefont {E.}~\bibnamefont
  {Kesteloot}},\ and\ \bibinfo {author} {\bibfnamefont {P.}~\bibnamefont {{Van
  Leuven}}},\ }\href
  {https://doi.org/https://doi.org/10.1016/0009-2614(88)80380-4} {\bibfield
  {journal} {\bibinfo  {journal} {Chemical Physics Letters}\ }\textbf {\bibinfo
  {volume} {149}},\ \bibinfo {pages} {547} (\bibinfo {year}
  {1988})}\BibitemShut {NoStop}%
\bibitem [{\citenamefont {Wecker}\ \emph {et~al.}(2015)\citenamefont {Wecker},
  \citenamefont {Hastings},\ and\ \citenamefont {Troyer}}]{VHA1}%
  \BibitemOpen
  \bibfield  {author} {\bibinfo {author} {\bibfnamefont {D.}~\bibnamefont
  {Wecker}}, \bibinfo {author} {\bibfnamefont {M.~B.}\ \bibnamefont
  {Hastings}},\ and\ \bibinfo {author} {\bibfnamefont {M.}~\bibnamefont
  {Troyer}},\ }\href {https://doi.org/10.1103/PhysRevA.92.042303} {\bibfield
  {journal} {\bibinfo  {journal} {Phys. Rev. A}\ }\textbf {\bibinfo {volume}
  {92}},\ \bibinfo {pages} {042303} (\bibinfo {year} {2015})}\BibitemShut
  {NoStop}%
\bibitem [{\citenamefont {Reiner}\ \emph {et~al.}(2019)\citenamefont {Reiner},
  \citenamefont {Wilhelm-Mauch}, \citenamefont {Sch{\"o}n},\ and\ \citenamefont
  {Marthaler}}]{VHA2}%
  \BibitemOpen
  \bibfield  {author} {\bibinfo {author} {\bibfnamefont {J.-M.}\ \bibnamefont
  {Reiner}}, \bibinfo {author} {\bibfnamefont {F.}~\bibnamefont
  {Wilhelm-Mauch}}, \bibinfo {author} {\bibfnamefont {G.}~\bibnamefont
  {Sch{\"o}n}},\ and\ \bibinfo {author} {\bibfnamefont {M.}~\bibnamefont
  {Marthaler}},\ }\href@noop {} {\bibfield  {journal} {\bibinfo  {journal}
  {Quantum Sci. Technol.}\ }\textbf {\bibinfo {volume} {4}},\ \bibinfo {pages}
  {035005} (\bibinfo {year} {2019})}\BibitemShut {NoStop}%
\bibitem [{\citenamefont {Luongo}(2020)}]{quantumalgorithms}%
  \BibitemOpen
  \bibfield  {author} {\bibinfo {author} {\bibfnamefont {A.}~\bibnamefont
  {Luongo}},\ }\href {https://quantumalgorithms.org} {\bibinfo {title} {Quantum
  algorithms for data analysis}} (\bibinfo {year} {2020})\BibitemShut {NoStop}%
\bibitem [{\citenamefont {Younis}\ \emph {et~al.}(2021)\citenamefont {Younis},
  \citenamefont {Iancu}, \citenamefont {Lavrijsen}, \citenamefont {Davis},
  \citenamefont {Smith},\ and\ \citenamefont {USDOE}}]{BQSKit}%
  \BibitemOpen
  \bibfield  {author} {\bibinfo {author} {\bibfnamefont {E.}~\bibnamefont
  {Younis}}, \bibinfo {author} {\bibfnamefont {C.~C.}\ \bibnamefont {Iancu}},
  \bibinfo {author} {\bibfnamefont {W.}~\bibnamefont {Lavrijsen}}, \bibinfo
  {author} {\bibfnamefont {M.}~\bibnamefont {Davis}}, \bibinfo {author}
  {\bibfnamefont {E.}~\bibnamefont {Smith}},\ and\ \bibinfo {author}
  {\bibnamefont {USDOE}},\ }\href {https://doi.org/10.11578/dc.20210603.2}
  {\bibinfo {title} {Berkeley quantum synthesis toolkit (bqskit) v1}} (\bibinfo
  {year} {2021})\BibitemShut {NoStop}%
\bibitem [{\citenamefont {Wigner}(1932)}]{wigner}%
  \BibitemOpen
  \bibfield  {author} {\bibinfo {author} {\bibfnamefont {E.}~\bibnamefont
  {Wigner}},\ }\href {https://doi.org/10.1103/PhysRev.40.749} {\bibfield
  {journal} {\bibinfo  {journal} {Phys. Rev.}\ }\textbf {\bibinfo {volume}
  {40}},\ \bibinfo {pages} {749} (\bibinfo {year} {1932})}\BibitemShut
  {NoStop}%
\bibitem [{\citenamefont {Johansson}\ \emph {et~al.}(2013)\citenamefont
  {Johansson}, \citenamefont {Nation},\ and\ \citenamefont {Nori}}]{qutip}%
  \BibitemOpen
  \bibfield  {author} {\bibinfo {author} {\bibfnamefont {J.}~\bibnamefont
  {Johansson}}, \bibinfo {author} {\bibfnamefont {P.}~\bibnamefont {Nation}},\
  and\ \bibinfo {author} {\bibfnamefont {F.}~\bibnamefont {Nori}},\ }\href
  {https://doi.org/10.1016/j.cpc.2012.11.019} {\bibfield  {journal} {\bibinfo
  {journal} {Computer Physics Communications}\ }\textbf {\bibinfo {volume}
  {184}},\ \bibinfo {pages} {1234} (\bibinfo {year} {2013})}\BibitemShut
  {NoStop}%
\bibitem [{\citenamefont {Guo}\ \emph {et~al.}(2023)\citenamefont {Guo},
  \citenamefont {Lou}, \citenamefont {Yu}, \citenamefont {Li}, \citenamefont
  {Fang}, \citenamefont {Liu}, \citenamefont {Long}, \citenamefont {Ying},\
  and\ \citenamefont {Ying}}]{10129229}%
  \BibitemOpen
  \bibfield  {author} {\bibinfo {author} {\bibfnamefont {J.}~\bibnamefont
  {Guo}}, \bibinfo {author} {\bibfnamefont {H.}~\bibnamefont {Lou}}, \bibinfo
  {author} {\bibfnamefont {J.}~\bibnamefont {Yu}}, \bibinfo {author}
  {\bibfnamefont {R.}~\bibnamefont {Li}}, \bibinfo {author} {\bibfnamefont
  {W.}~\bibnamefont {Fang}}, \bibinfo {author} {\bibfnamefont {J.}~\bibnamefont
  {Liu}}, \bibinfo {author} {\bibfnamefont {P.}~\bibnamefont {Long}}, \bibinfo
  {author} {\bibfnamefont {S.}~\bibnamefont {Ying}},\ and\ \bibinfo {author}
  {\bibfnamefont {M.}~\bibnamefont {Ying}},\ }\href
  {https://doi.org/10.1109/TQE.2023.3275868} {\bibfield  {journal} {\bibinfo
  {journal} {IEEE Transactions on Quantum Engineering}\ }\textbf {\bibinfo
  {volume} {4}},\ \bibinfo {pages} {1} (\bibinfo {year} {2023})}\BibitemShut
  {NoStop}%
\bibitem [{\citenamefont {Mohan}\ \emph {et~al.}(2025)\citenamefont {Mohan},
  \citenamefont {Bhowmick}, \citenamefont {Kumar},\ and\ \citenamefont
  {Chaurasiya}}]{Mohan_2025}%
  \BibitemOpen
  \bibfield  {author} {\bibinfo {author} {\bibfnamefont {N.~K.}\ \bibnamefont
  {Mohan}}, \bibinfo {author} {\bibfnamefont {R.}~\bibnamefont {Bhowmick}},
  \bibinfo {author} {\bibfnamefont {D.}~\bibnamefont {Kumar}},\ and\ \bibinfo
  {author} {\bibfnamefont {R.}~\bibnamefont {Chaurasiya}},\ }\href
  {https://doi.org/10.1088/1402-4896/adb7a1} {\bibfield  {journal} {\bibinfo
  {journal} {Physica Scripta}\ }\textbf {\bibinfo {volume} {100}},\ \bibinfo
  {pages} {035118} (\bibinfo {year} {2025})}\BibitemShut {NoStop}%
\end{thebibliography}%
\selectlanguage{american}%

\end{document}